# Excitation of vortices using linear and nonlinear magnetostatic waves

A.D. Boardman*, Yu.G. Rapoport**, V.V. Grimalsky***, B.A. Ivanov****,

S.V. Koshevaya*****, L. Velasco*, C.E. Zaspel******

*Joule Physics Laboratory, University of Salford, Salford, Greater Manchester,M5 4WT, UK, email: a.d.boardman@salford.ac.uk

**Kyiv National University, Physical Faculty, prospect Glushkov, 6, 03022, Kyiv, Ukraine, e-mail: laser@i.kiev.ua

***National Institute for Astrophysics, Optics, and Electronics, Puebla 72000, Pue., Mexico

****Resesarch Inst. of metallophysics, Ukrainian Acad. Of Sci., Kiev, Ukraine

*****Autonomous University of Morelos (UAEM), CIICAp, Faculty of Chemistry, Av. Universidad, 1001, Z.P. 62210, Cuernavaca, Mor., Mexico

****** University of Montana/Western, Dillon, MT 59725 USA

It is shown that stationary vortex structures can be excited in a ferrite film. This is the first proposal for creating vortex structures in the important cm and mm wavelength ranges. It is shown that both linear and nonlinear structures can be excited using a three-beam interaction created with circular antennae. These give rise to a special phase distribution created by linear and nonlinear mixing. An interesting set of three clockwise rotating vortices joined by one counter-rotating one presents itself in the linear regime: a scenario that is only qualitatively changed by the onset of nonlinearity. It is pointed out that control of the vortex structure, through parametric coupling, based upon a microwave resonator, is possible and that there are many interesting possibilities for applications.

……………………………………………………………………………………

## 1. INTRODUCTION

The recognition that vortices are rather important manifestations of optical phase singularities has opened up a new frontier in optics. These 'swirling' entities can, in principle, be used to carry data and point to a new era for communications. While electromagnetic vortices are attractive for modern optical applications, it should be noted that vortices, in general, are a common phenomenon and have been observed both in water and the atmosphere for hundreds of years. They are a common occurrence in plasma and atmospheric science, and striking examples include the behavior of Rossby [1] and Alfven waves [2]. Even so, it is only in recent times that the wave field structure has been understood properly, from an electromagnetic point of view, and the full generic mathematical properties have been developed. Experimentally, vortices in liquids [3], space plasmas [4] and the atmosphere [5] have been thoroughly investigated but it is the recent studies, especially in nonlinear optics, that have stimulated an almost explosive development, both in theoretical and experimental contributions [6, 7]. This work builds upon the pioneering work of Nye and Berry [8].

The successes of optical vortex theory and experiment have an importance that goes far beyond the boundaries of high frequency optics because the generic ideas ought to be applicable to magnetostatic waves and should impact upon quasi-optical microwave and mm-wave systems. The latter are of increasing importance in surveillance utilizations, so any new ideas in this area are likely to generate many applications. Up to now, however, no arrangements have been designed to excite vortices in the microwave or millimeter frequency range, even though the creation of a "vortex antenna" in the microwave and mm- ranges would be very important.

This paper addresses how this important step can be taken, by investigating phase singularities in forward volume magnetostatic waves [9] (FVMSW). The starting point is the famous Nye and Berry [8] definition, which states that a phase singularity in space occurs where the real and the imaginary parts of the wave field vanish simultaneously i.e. for a scalar wave field, represented by the complex function F, a phase singularity whenever

$$F_{\mathbf{r}} = 0; \quad F_i = 0\,. \tag{1a}$$

Here $F_{\mathbf{r}}$ and $F_i$ are the real and imaginary parts, respectively, and the phase $\Phi$ is defined through

$$\tan(\Phi) = \frac{F_i}{F_r}\ . \tag{1b}$$

Obviously, $\Phi$ is indeterminate, whenever equation (1a) is satisfied; hence the term 'singularity'. Actually, this singularity may be associated with either an edge or a screw dislocation and it is the latter that is characterizes vortices, which will be addressed here. The presence of a vortex means that the line integral of the spatial gradient of the phase, taken around the vortex center, or line, is some multiple of π i.e.

$$\oint (\nabla\Phi) dl = 2n\pi \qquad (2a)$$

in which $n = 1,2,...$ is a positive integer, proportional to what has become known as the "topological charge". If $n = 0$ then a dipole vortex structure is said to exist.

The Poynting vector for the magnetostatic waves used in this paper is [10]

$$\vec{P} = \frac{c}{8\pi} \mathrm{Re}(\varphi \frac{\partial \vec{b}^*}{\partial t}) \qquad (2b)$$

where $\varphi$ and $\vec{b}$ are the magnetostatic potential and the magnetic induction, respectively. If a singularity like a vortex is present then the Poynting vector will rotate around the "vortex center", which is a point in the two-dimensional case and an axis in the three-dimensional case. The surface of constant phase associated with a screw dislocation is helicoidal, so that the Poynting vector, which is directed along the normal, spirals about the propagation direction.

In the magnetostatic regime of a ferrite material

$$\boldsymbol{h} = -\vec{\nabla}\varphi . \qquad (2c)$$

where $\boldsymbol{h}$ is the magnetic field and there is a potential carried by a magnetostatic wave of the form

$$\{ \varphi = |\varphi| \exp[i\Phi] \qquad (2d)$$

where $\Phi$ is the phase and the amplitudeis $\varphi$ .

The scalar function $\varphi$ and $h_Z$ , the component of magnetic field normal to the surface of the ferrite film, sketched in Fig. 1, both satisfy conditions (1), (2), for $n \neq 0$ . On the other hand, the tangential components of the magnetic field $\boldsymbol{h}$, lying in the $OXY$ plane of the ferrite film satisfy conditions (1), (2) with $n = 0$.

## 2. CIRCULAR ANTENNA EXCITATION OF LINEAR FORWARD VOLUME MAGNETOSTATIC WAVES

The geometry for forward volume magnetostatic waves (FVMSW) and a sketch of the excitation arrangement are shown in Figs. 1, 2. A linear FVMSW propagates in a ferrite film along the direction $Y$ , normal to an applied magnetic field that is directed along the Z-axis. The magnetostatic potential and field components are proportional to exp(i(ωt)),where $\omega$ is the angular frequency. The field structure and dispersion relation are determined through the equation [9, 10]

$$div\vec{b} = [\mu(\varphi_{XX} + \varphi_{YY}) + \varphi_{ZZ}] = 0$$

(3)

where

$$b_X = -(\mu\varphi_X - i\mu_a\varphi_Y),\ b_Y = -(\mu\varphi_Y - i\mu_a\varphi_X),\ b_Z = \varphi_Z, \qquad (4a)$$

inside the ferrite film, and

$$div\,\vec{b} = -(\varphi_{XX} + \varphi_{YY} + \varphi_{ZZ}) \qquad (4b)$$

outside the ferrite film. The quantities $\mu$ and $\mu_a$ are components of the permittivity tensor,

$$\mu = \frac{\omega_\perp{}^2 - \omega^2}{\omega_H{}^2 - \omega^2},\ \mu_a = \frac{\omega\,\omega_M}{\omega_H{}^2 - \omega^2}, \qquad (4c)$$

in terms of the definitions $\omega_H = |\gamma|(H_0 - 4\pi M_0), \omega_M = |\gamma| 4\pi M_0$, and $\omega_\perp^2 = \omega_H(\omega_H + \omega_M)$, in which $H_0$ and $4\pi M_0$ are the bias externally applied magnetic field and the saturation magnetization of the ferrite film, respectively. $\gamma$ is called the gyromagnetic ratio. The tangential components of magnetic field $\varphi_{X,Y}$ and the normal component of magnetic induction, $b_Z$ are continuous at the ferrite film boundaries. The latter are located at $Z = \pm L/2$, and the shape of the fundamental mode of the magnetostatic potential [9] is

$$\varphi = \begin{cases} \dfrac{\cos(\sqrt{-\mu}\,|k|\,Z)}{\cos(\sqrt{-\mu}\,|k|\,L/2)}, & |Z| \le L/2 \\ e^{-|k|Z}, & |Z| > L/2 \end{cases} F(X,Y)e^{i\omega t} \equiv f_z(z)F(X,Y)e^{i\omega t}, \qquad (5a)$$

where $k$ is determined from the dispersion equation

$$\tan(\sqrt{-\mu}kL/2) = \frac{1}{\sqrt{-\mu}} \qquad (5b)$$

Hence the equation for $F$ is

$$\frac{\partial^2 F}{\partial X^2} + \frac{\partial^2 F}{\partial Y^2} + k^2 F = 0. \qquad (5c)$$

For an excitation of a forward volume magnetostatic wave by the kind of circular antennae shown in Fig.2, the solutions are complicated by the cylindrical geometry.

In the Fig.2, circular antennae are placed at $\frac{2\pi}{3}$ intervals around the center at O. The distances from the center to the chords of these arcs are labeled $OO_i$ (i =1,2,3). The chords are segments of the triangle ABC. When excitation of the FVMSWs by arc antennas is considered, the geometrical optics is used to "transform or recalculate" the boundary conditions from the arcs to "effective straight line antennae" (see the end of the Sect. 3). The length of the section $X_1X_0$ is found as follows. $|O_1X_1|= x_1$, where $x_1$ is the coordinate of the point $X_1$ at the section $O_1C_1$. Radius of the arc $B_1C_1$ is $r = |OX_0| = |OB_1|; |OO_1| == \sqrt{r^2 - \Delta^2}$;

; $|OX_1| = \sqrt{|OO_1|^2 + |O_1X_1|^2} = \sqrt{|OO_1|^2 + {x_1}^2} = \sqrt{r^2 - \Delta^2 + {x_1}^2}$ . Therefore

$$|X_1X_0| = |OX_0| - |OX_1| = r - \sqrt{r^2 - \Delta^2 + {x_1}^2} \approx \frac{\Delta^2 - {x_1}^2}{r} . \quad (6)$$

To be specific, only solutions with finite values of magnetostatic potential in the center of the circular antenna need to be considered. Hence, $\varphi$ and $F(r,\theta)$ are finite, at $r = 0$, where $(r,\theta)$ are cylindrical coordinates, with the origin coinciding with the center of the circular microstrip antenna. The Poynting vector in this system of coordinates has the components:

$$P_r = -\frac{c}{8\pi}\mathrm{Re}[i\omega\varphi(i\mu_a \frac{1}{r}\frac{\partial\varphi^*}{\partial\theta} + \mu\frac{\partial\varphi^*}{\partial r})], P_\theta = -\frac{c}{8\pi}\mathrm{Re}[i\omega\varphi(-i\mu_a \frac{\partial\varphi^*}{\partial r} + \mu\frac{1}{r}\frac{\partial\varphi^*}{\partial\theta})] \quad (7)$$

If the radius of the microstrip antenna is $r_0$, then the magnetostatic potential is determined by the additional boundary condition

$$F(r_0,\theta) = Ce^{i\theta}, \quad (8a)$$

where $C$ is a real constant. The solution of eq.(5c) is, therefore,

$$F(r,\theta) = Ce^{i\theta}\frac{J_1(kr)}{J_1(kr_0)} . \quad (9)$$

from which it is clear that $F(0,\theta) = 0$ and a property associated with a vortex structure is guaranteed for FVMSW excited by the circular antenna. The contour integration of the phase along a circular path around the phase defect at $r = 0$, leads to $\oint_L d\Phi = \int_{0^+}^{2\pi} d\theta = 2\pi$, with the conclusion that $n = 1$ and that the topological charge is 1. Although the field is determined by means of a single, scalar, magnetic potential function, it is also interesting to approach the investigation of the topological structure through the components of the magnetic field vector . The latter, for points on a circle of the radius $r = \rho$, is

$$\vec{h}\,|_{Z=L/2} = -\vec{\nabla}\varphi\,|_{Z=L/2} = \vec{e}_Z h_Z + \vec{e}_r h_r + \vec{e}_\varphi \vec{h}_\varphi = \\ \vec{e}_Z kCe^{i\theta}\,\frac{J_1(k\rho)}{J_1(kr_0)} + \vec{e}_r kC^{i\theta}\,\frac{J_1'(k\rho)}{J_1(kr_0)} + \vec{e}_\theta\,\frac{1}{\rho}Ce^{i\theta}\,\frac{J_1(k\rho)}{J_1(kr_0)} \tag{10}$$

where the common multiplier $e^{i\omega t}$ is omitted.

It is interesting that, although the real and imaginary parts of field components $h_Z, h_\theta$ vanish at the center of the circle and lead to the same topological structure, the component $h_r$ does not satisfy the condition of vanishing at the origin. Further confirmation of the presence of a vortex structure comes from the behavior of the Poynting vector, which rotates around the center. In fact, the only non-zero component of the Poynting vector is

$$P_\theta = \frac{c}{8\pi}|C|^2\, I_z\,\frac{J_1(kr)}{J_0{}^2(kr_0)}[\frac{\mu}{r}J_1(kr) - \mu_a kJ_1'(kr)], \tag{11}$$

where $I_z = \int_{-\infty}^{\infty} f_z{}^2(z)dz$, and $f_z(z)$ comes from (5a).

In summary, for field distributions associated with a circular antenna structures with the kind of phase defects expected for vortex creation exist for the magnetostatic potential and the field components $h_Z, h_\theta$, with a topological charge $n = 1$, but for the field component $h_r$ they exist for $n = 0$.

From a practical point of view it is easier to use several short plane, or arc antennae, placed along the corresponding circle, instead of deploying an entire circular antenna. In this paper two cases are developed, one consisting of three plane antennae and the other of three arc-shaped ones. All are located along the same circle and so the analysis rests upon a three-beam interaction. To simulate the phase distribution, a phase shift of $2\pi/3$ between the neighboring antennae will be used. The first case, addressed in the next section, is a simplified model of a three-linear plane wave interaction.

## 3. A LINEAR PHASE DEFECT STRUCTURE INDUCED BY THREE PLANE WAVES.

The creation of a linear scalar vortex structure, through the interaction of three plane waves in a *bulk crystal*, has already been considered in the optical domain [11], but here a vector structure is considered using three forward volume magnetostatic waves in a ferrite thin film. As stated earlier, the directions of propagation of the three plane waves are displaced at angles $2\pi/3$ , with respect to each other. In addition, they will also be assumed to have amplitudes $A_1 = A_2 = A_3 = A$.

A phase defect of the magnetostatic potential $\varphi$ is placed at the "center of interaction", which is the point O in the Fig. 2. If it is assumed that the distances from all three antennas (shown as sections of solid lines $B_1O_1C_1$, $A_2O_2C_2$ and $A_3O_3C_3$ in Fig. 2) to the center at O are all equal, then $|O_1O|=|O_2O|=|O_3O|= r_0$. The initial phases of the potential at the antennae 1, 2 and 3 are $\Phi_1, \Phi_2$ and $\Phi_3$, respectively. Now consider the magnetostatic potential at points on the circle $r = \rho$ surrounding the center O, and set

$$k\rho << 1 \tag{12a}$$

Given that the polar axis is directed from point O, the center of the interaction, towards the center of the first antenna of Fig. 2, the potential at some point P, with polar coordinates $\rho, \theta$, is

$$\varphi = Ae^{-ikr_0}e^{i\omega t}[e^{i\Phi_1}e^{ik\rho\cos\theta} + e^{i\Phi_2}e^{ik\rho\cos(\theta-2\pi/3)} + e^{i\Phi_3}e^{ik\rho\cos(\theta-4\pi/3)}]f_z(z) \tag{12b}$$

Putting $\Phi_1 = 0, \Phi_2 = 2\pi/3, \Phi_3 = 4\pi/3$ the term in the square brackets of (12) plays the role of an "effective complex envelope", $\tilde{A}$, where

$$\tilde{A} = e^{i\Phi_1}e^{ik\rho\cos\theta} + e^{i\Phi_2}e^{ik\rho\cos(\theta-2\pi/3)} + e^{i\Phi_3}e^{ik\rho\cos(\theta-4\pi/3)} = \tilde{A}' + i\tilde{A}'', \tag{13}$$

and, under the condition (12b),

$$\tilde{A}' \approx -(3/2)k\rho\sin\theta, \tilde{A}'' \approx (3/2)k\rho\cos\theta, \tag{14a}$$

which leads to the equivalent representation

$$\tilde{A} = (3/2)k\rho e^{i(\theta+\pi/2)}, \ \varphi = Ae^{-ikr_0}e^{i\omega t}\tilde{A}f_z(z). \tag{14b}$$

Obviously, the singularity condition (1a) is satisfied at the center of the circle. The phase $\Phi$ of complex amplitude $\tilde{A}$ is introduced through

$$\tan\Phi = \frac{\tilde{A}''}{\tilde{A}'}, \ d\Phi = \frac{\tilde{A}'d\tilde{A}'' - \tilde{A}''d\tilde{A}'}{\tilde{A}'^2 + \tilde{A}''^2}. \tag{15}$$

If the point P is on the circle of radius $\rho$, then $d\Phi = d\theta$ and the phase gradient integral (2a) will yield the "topological charge" associated with a magnetostatic potential phase defect. Hence it is clear how a phase defect structure of a magnetostatic potential can be created. What is needed are three, equal amplitude, interacting plane waves, moving under the angle restraint that they have normals that are inclined at $2\pi/3$, with respect to each other. Naturally this conclusion rests upon choosing appropriate initial amplitudes. It is straightforward to show that where $k\rho << 1$ the components of the Poynting vector are

$$P_r = 0,\ P_\theta = \frac{c}{8\pi} I_z \frac{9}{4} |A|^2 \omega k^2 r(\mu_a - \mu), \tag{16}$$

which demonstrates that the Poynting vector rotates around the center, as required for vortex-like structures to make their appearance. Some simulations to illustrate this case are shown in figure 3.

## 4. ARC ANTENNA CREATION OF LINEAR AND NONLINEAR STRUCTURES WITH PHASE DEFECTS

The "antenna structure" is placed on the surface of the kind of normally magnetized ferrite film shown in the Fig. 1. As before, the structure consists of microstrip antennae generating three interacting stationary forward volume magnetostatic waves. This time, however, the antennae are the arcs of a circle that are labeled 1, 2 and 3 in Fig 2. The boundary conditions are transformed in this case to make "effective antennae" in the form of straight line segments $B_1O_1C_1$, $A_2O_2C_2$ and $A_3O_3C_3$ shown in Fig. 2. FVMSW beams interact principally in the region 4, but diffraction and possible nonlinearity change the simple "geometrical optics" picture of beam propagation. The interaction regions shown in the Fig. 2 are shown to provide a simplified picture only. The interaction of three FVMSW beams in dimensionless form is modeled by the coupled equations

$$\frac{\partial U_j}{\partial y_j} + ig\frac{\partial^2 U_j}{dx_j^{\,2}} + iN(|U_j|^2 + 2\sum_{l\neq j}|U_l|^2)U_j + \gamma U_j = 0, \tag{17}$$

where $j,l = 1,2,3$, $U_j$ is a dimensionless complex amplitude of the magnetic potential and $N$ is dimensionless nonlinear coefficient:

$$\varphi = (1/2)U_{dj} f_z(Z)\exp[i(\omega t - ky_j)] + c.c, U_{dj} = U_j U_0 \tag{18a}$$

where $U_{dj}$ are the true dimensional amplitudes of j-th beam and $U_0$ is an amplitude used for normalization. $y_j$ is the direction of j-th beam propagation in the corresponding coordinate (see Figs. 1, 2), and $f_z(Z)$ is the transverse distribution function in the ferrite film (see also eq. (5a) ).

$$g = \frac{1}{2l_0 k}, N = N_d U^2{}_0 l_0, \gamma = \gamma_d l_0$$

(18b)

are the respective dimensionless parameters of diffraction, self-interaction, and loss, in which $N_d = (\frac{\partial k}{\partial |U_d|^2})_{\omega=const} = -\frac{1}{V_g}(\frac{\partial \omega}{\partial |U_d|^2})_{k=const}$ and $\gamma_d$ are the actual nonlinear coefficient and coefficient of loss respectively. Typically,

$$l_0 = 1cm, \gamma = 0.2, N = 1 \qquad (18c)$$

so these have been selected for the calculations reported here. In addition, the numerical simulations are based upon a ferrite film of thickness $L = 10\mu m$, and $|BC| = a = 1cm$. The frequencies and wave numbers of each of the three interacting FWMSW are identical and are set equal to $\omega \approx \omega_H \approx \omega_M \sim 3 \cdot 10^{10} s^{-1}$, $k \approx 150 cm^{-1}$. The nonlinear coefficient, $N_d$, for an exchange-free nonlinear Schrodinger envelope equation can be determined using expansion in series by small amplitudes of the nonlinear dispersion equation [12] or by means of bilinear relationships analogical to the energy conservation law [13]. Estimates show that for $N = 1$, $l_0 = 1cm$, the normalizing amplitude is such as $kU_0 \sim 2$ Oe, or $U_0 \sim 1.3 \cdot 10^{-2}\ Oe \cdot cm$; the value of $N_d$ can be found from the relation (18b) with a given $N$ and $l_0$ (see also (18c) ) and the estimated value of $U_0$.

The beams converge more towards the center of the interaction, taking into account some possible diffraction, through the bending of the antennae into arcs. The boundary conditions are introduced by transforming the conditions from the actual arc antennae into those that pertain to "effective straight line antennae". As a result of this strategy, the following analogous boundary conditions are used for beams $i = 1, 2, 3$, using the first one as a reference beam,

$$U_i(x_i, y_i) = U_{i0} \exp[-((x_i - x_{i0})/x_0)^2] \exp[i\Phi_i + i\psi(x_i)]. \qquad (19)$$

Here $U_{i0}$ is the maximum amplitude of the $i$th beam, the initial phases $\Phi_i$ are determined by the relationships defined earlier, all the beam widths are $x_0$, $x_{i0}$ are the positions of the beam centers, relative to the central points of the antennae and the the phases $\psi(x_i)$ arise because of the transformation from the arcs to the corresponding straight segments e.g., for beam 1, the transition is from arc 1 to the straight segment $B_1C_1$, , which is illustrated in Fig. 2 for the case when $x_{i0} = 0$. Hereafter beams with $x_{i0} = 0$ and $x_{i0} \neq 0$ are called "unshifted" and "shifted" beams respectively. To find the phases $\psi(x_i)$, consider, for the sake of definiteness, the beam in Fig. 2, for which $|B_1C_1| = 2\Delta << r = \frac{a\sqrt{3}}{6}$, where the radius of the circle is $r = |OB_1|$ and $\Delta$ is equal approximately to the half-length of the FVMSW antenna. An application of simple ray optics and using eq. (6) then leads to the result (see also captions to the Fig. 2)

$$\psi(x_1) = -ik\,|X_1X_0| = r - \sqrt{r^2 - \Delta^2 + x_1^{\,2}} \approx \frac{\Delta^2 - x_1^{\,2}}{r} \tag{20}$$

## 4.SPATIAL PATTERNS OF EXCITED LINEAR STRUCTURES

Setting $N = 0$ yields linear structures and these are shown in Figs. 3, 4 and 5. These cover the following cases. (a) no focusing or shifting of the beam for which $\exp[i\psi(x_i)] = 1$ and $x_{i0} = 0$ (b) focusing without shifting of the beam, for which $\exp[i\psi(x_i)] \neq 1$ and $x_{i0} = 0$ (c) $\exp[i\psi(x_i)] \neq 1$ and $x_{i0} \neq 0$, which means that there is both focusing and beam shifting. Note also that Figs 4, 5, 6 correspond to the interaction of three magnetostatic waves with equal absolute amplitude values. Figs. 3a, 3a, 3a, Figs. 3b, 3b, 3b, and Figs. 3c, 3c, 5c, respectively, show the spatial distributions of $\mathrm{Re}\,\varphi(x, y)$, $|\varphi(x, y)|^2$, and a vector proportional to the Poynting vector. If required the coefficient of proportionality can be determined in each particular case and has only been selected from the point of view of clarity and quality of presentation of the essential features. $h_z \sim \partial\varphi / \partial z \sim F(x, y)\frac{\partial f_z(z)}{\partial z}$, $\mathrm{Re}\,\varphi(x, y)$ and $|\varphi(x, y)|^2$ have the following physical interpretations. To within an accuracy of some multiplier, they are $\text{Re}\,\varphi(x.z)\,|_{z=L/2} \sim \text{Re}\,F(x,y) \sim h_z\,|_{z=L/2}$, $|\,\varphi(x,z)\,|^2{}_{z=L/2} \sim |\,h_z\,|^2{}_{z=L/2}$. Without focusing (see Fig. 3), in the vicinity of the center, a magnetostatic wave beam, with a finite width equal to $x_0$, is approximately a plane wave. It can be seen from Figs. 3a,b that, as a result of a linear interaction among the three waves, a structure with a quasi-periodic set of "singular points" is generated. Figs. 4 a, b show clearly the effect of focusing in the central region of the three beams excited by arc antennae. In this case, singular quasi-periodic structures for $\text{Re}\,\varphi(x,y)$ and $|\,\varphi(x,y)\,|^2$ are also generated in the central region of the ferrite film. Quasi-periodic structures for $\text{Re}\,\varphi(x,y)$ and $|\,\varphi(x,y)\,|^2$ are absent, however, when the three focused beams have centers displaced, with respect to the center of the interaction i.e. $x_{i0} \neq 0$, as can be appreciated from Figs 5 a,b. Quasi-periodic vortex structures around O with (dimensionless) coordinates $x = 0, y = 0.287$, or (dimensional) $x = 0, y = a/(2\sqrt{3})$) ,for undisplaced beams are shown in Figs. 3c [no focusing] and 4c [focusing]. The formation of a system of vortices, with right and left directions of rotation of the Poynting vector, is obvious from these Figures. The value of the magnetostatic potential $\varphi$ was found at O by a direct numerical calculation. It is zero in both the linear (Figs. 3-5) and the nonlinear cases, when the absolute values of amplitudes of all three interacting waves are equal to each other.

In the case of three beams with shifted centers ($x_{i0} \neq 0$) a symmetrical structure of "right-" and "left-" rotating vortices is formed around O as seen in Fig. 5c. In the linear case, for equal amplitudes of the interacting beams, the unit value of topological charge was obtained by direct numerical computation. In the general nonlinear wave case, however, a "geometrical" method is more convenient. For both focusing and nonlinear beams, direct computation of the phase integral is not very accurate because the accuracy is restricted by the value $k\sqrt{dx^2 + dy^2}$, which is of order of 0.15. Even though this is an adequate limitation for the calculation of the potential envelope function, within the usual slowly varying approximation, the inaccuracy of the phase integral is rather large. An effective alternative to direct computation of this integral, however, is to build a graph of the phase along some circumference. Taking into account that each jump of phase is equal to $\pi$, the number of "jumps" in the phase along this circumference then yields a value of "topological charge". Fig.6a shows the spatial distribution of the phase in the vicinity of the origin for linearly interacting focusing beams with equal absolute values of amplitudes. Two phase jumps, each approximately equal to $\pi$, can be clearly seen, so that the topological vortex charge" of the corresponding vortex is equal to 1. Fig. 6b shows the same phase distribution, but over a larger part of interaction region.

Note that the "motion" of magnetostatic potential in the vicinity of the center can be described as a rotation. There is a localization of the magnetostatic potential along Z direction (in the region of the film) which is described by function included in the relationship (5a). The magnetostatic potential inside the ferrite film is the result of interference of two waves counter propagating and forming a "standing" wave in the Z direction. The phase defects for non-focused and focused interacting beams, respectively, can be called "arrangements of quasiscrew phase dislocations" or "a set of quasiscrew vortices" in a ferrite film. The term "quasiscrew" is meant to convey the fact that a real screw is replaced in the ferrite film by a rotation. If three linear (optical) plane waves in an unbounded media have wave vectors in the same plane, and equal amplitudes, wavenumbers and frequencies, a periodic set of screw dislocations occurs. The symmetry of the wave structure with a set of phase defects that are vortices is interesting. For the geometry depicted in Fig. 2, using three waves with equal amplitudes, wavenumbers and frequencies and initial phases equal to $\Phi_1 = 0, \Phi_2 = 2\pi/3, \Phi_3 = 4\pi/3$, the total magnetostatic potential satisfies the following relationship :

$$\varphi(\theta_0 + 2\pi/3) = \varphi(\theta_0)e^{i2\pi/3}. \qquad (21a)$$

where $\theta_0$ is a polar angle and the relationship describes the "macroscopic symmetry" of the structure considered. In distinction to this, circular symmetry is evident only in the neighbourhood of the "vortex axis". For three identical interacting, lossless, linear, plane waves with equal amplitudes the analysis works well within the vicinity of any phase defect. If losses, diffraction, nonlinearity or focusing are present, the amplitude of each of the waves will depend on the coordinates, and eq. (14a) will be valid only for the vicinity of the geometrical center of the structure. In the latter case, in the other points of the phase defects shown in Figs. 3a,b and 4 a,b, eq. (14a) will be valid only approximately, because supposition that all the wave amplitudes are equal to each other, will not be true in the vicinity of other phase defect points. At the same time, because the amplitudes of the waves change slowly in time and space it can be said, at least for phase defect seen in Figs. 3a,b, 4a,b, that being in the neighbourhood of the center gives them validity. Therefore, even for lossy and dispersive (Fig. 3a,b), focused (Fig. 4a,b), or nonlinear (see Figs. 8, 9 below) beams, the structures can be treated as a "set of quasiscrew dislocations". As can be seen from eq. (14a), and also from the way that the phase is determined (see eqs. (1b), (15) )

$$\Phi(\theta + \pi) = \Phi(\theta) \quad (21b)$$

Fig. 7 that shows a small fragment of the phase distribution with equal phase contours, for three focused linear beams with equal amplitudes (Figs. 6a,b) in the close vicinity of the central point and demonstrates the above phase periodicity. Note that Fig. 7 is a, small fragment of the Fig. 6 so it reflects the same structural symmetry as Fig. 6. Equation (21b) characterizes the "microscopic" symmetry of the structure in the neighbourhood of the phase defect point(s) while (21a) characterizes the "macroscopic" symmetry of the structure as a whole, relative to the geometrical center.

## 5. INFLUENCE OF NONLINEARITY, UNEQUAL BEAM AMPLITUDES AND SHIFTED BEAM CENTERS

For a linear three-wave magnetostatic interaction, defined with a suitable relative phase shift, a rotating vortex structure has been established in the previous sections. The influence of nonlinearity upon the shape of this structure and also the "vortex charge" will now be addressed, together with a study of the role of interacting beam asymmetry , when the moduli of their amplitudes are unequal. The work described here can be put into a broader context by noting that, in the paper [14] devoted to nonlinear singular optics, in particular, three-wave interaction was considered and the possibility of vortex charge "redistribution" was discussed.

Suppose, for simplicity, that only the first two polar modes are excited i.e., $F_{12}(r,\theta) = A_1(r)e^{i\theta} + A_2(r)e^{2i\theta}$, where values $A_{1,2}$ are real and $F_{12}(0,\theta) = 0$. and $F_{12}$ is a function describing qualitatively a vortex structure, with a center at $r = 0$. The outcomes of a direct calculation, of the integral of the phase gradient are $2\pi, 3\pi$ and $4\pi$, respectively, for $A_2 = 0, A_1 = A_2$ and $A_1 = 0$. A more general result can be easily obtained using $F_{12}(r,\theta) = A_1(r)F_1(\theta)F_2(r,\theta)$, where $F_1(\theta) = e^{i\theta}, F_2(r,\theta) = 1 + r_0(r)e^{i\theta}$ and $r_0(r) = A_2(r)/A_1(r)$. A simple geometrical arguments is enough to investigate the phase integrals. for $r_0 < 1, r_0 = 1$ and $r_0 > 1$.

The corresponding “vortex charges” are 1, 3/2 and 2 in these cases, so that if the ratio of amplitudes of second and main polar harmonics can be changed (linearly or nonlinearly), the “vortex charge” can also be changed. The transformations of topological charge during free-space propagation of a light wave, which is a combination of a Gaussian beam with a multiple charged optical vortex within a Gaussian envelope, were studied in [15]. In addition, it interesting to observe that the possibility of generating an optical vortex with a fractional charge after the diffraction of a linear optical beam on thin binary amplitude grating has already been discussed [16].

In the present calculations, only a rather low level of harmonics are excited, and they determine nonlinear coefficient needed for the envelope equation. The nonlinearity provides self- and cross- interactions of the main harmonic amplitudes, in the vicinity of the singular points. The higher harmonics are included only implicitly into the system of coupled envelope equations through the nonlinear coefficient because the amplitudes of harmonics are much smaller than the amplitudes of three main interacting beams. A small enough nonlinearity will not influence the charge of the generated vortex. The numerical calculations presented below confirm this conclusion.

Figs. 8-13 are generated for dimensionless magnetic field amplitudes $U_j \sim (2 \div 3)$; values that are necessary to provide the nonlinear effects described below. Note that the further increase of the input amplitudes may lead to the second-order spin-wave instability [17-19]. A possibility of exciting the exchange spin waves (with the same frequency as the incident beams) is not taken into account in the present paper and will be the subject of future work. The nonlinear calculations, presented here, show a qualitative influence of nonlinearity on vortex formation due to a three beam interaction. The influence of the nonlinearity upon the shape of the spatial structures can be seen in Figs. 8, 9, which are computed with equal absolute values of all nonlinear beam amplitudes. They show what happens for cases in which there is an absence and or presence of focusing. Comparing Figs. 8, 9 with Figs. 3a, 4a shows that the nonlinearity leads to nonlinear 2D diffraction. This causes broadening of the region of interaction and this is accentuated for the focused_beams, as can be appreciated by looking at Figs. 4a and 9. The difference between the linear and nonlinear structures, in the absence of focusing is rather less than in the presence

of focusing. Simultaneous asymmetry of the interacting beams brings out the most pronounced effect of the nonlinearity. As can be seen from the Figs. 4c and 5c, equal amplitude linear focused beams leads to a system of right- and left- rotating vortices {Should we use the word, vortices since it is more commonly seen in the literature?} that are symmetrical, regardless of whether they are shifted, or not. If the linear beams have only slightly different absolute amplitudes, almost symmetric vortex structures are maintained and are shown in Figs. 10 and 11. "Switching on" the nonlinearity, for the amplitude values used in the linear structures, changes the situation to that shown in Figs. 12 and 13. Here it can be seen that the vortex structure formed by unshifted beams, endowed with different absolute amplitudes, shifts only slightly "as a whole" due to nonlinearity. If the beam centers are shifted then nonlinearity causes a rather complicated rebuilding of the vortex structure topology, as seen by comparing Figs. 5c and 11 with the Fig. 13. Fig. 13 reveals that the nonlinearity causes a rotation of the structure as a whole, spatial shifting of the points with phase defects and a rebuilding that includes merging with a neighboring part of a structure. Overall the conclusion is that in the absence of nonlinearity, slightly different amplitudes for the interacting beams permits the symmetry to remain but nonlinearity causes it to disappear. Note that it is also probably possible to create vortices using structures containing layers with higher nonlinearity than pure YIG films possess e.g. "ferrite-paraelectric" structures [20, 21].

Using just the visual evidence, presented in the form of the Poynting vector and phase spatial distributions, no points have been discovered where, simultaneously, the complex amplitude of the magnetostatic potential is zero and the phase integral around this point is different from $2\pi$. In other words, for the linear, or nonlinear, interaction of three magnetostatic beams, under the condition that there is only an insignificant presence of higher harmonics, the "vortex charge" does not differ from 1 or –1.

## 7. DISCUSSION AND CONCLUSIONS

It has been shown that some stationary vortex structures could be excited experimentally in a ferrite film. One suggestion is that the vortex structure can be exposed by a Brillouin scattering method [22]. This is the first proposal of a method of creating vortex structures in the technically important cm and mm- ranges. It may use YIG, or hexaferrite material, but without using superconductivity, or an exchange interaction. The main requirement for an experimental arrangement would be to maintain the necessary difference of phases between the antennae exciting three interacting beams. In the future, nonstationary structures, and the combination of vortex excitation with parametric interaction will be considered.

On the basis of the present work linear and nonlinear vortex structures can be excited in a ferrite film, using a three-beam interaction with a $2\pi/3$ phase difference between the $2^{nd}$ and $1^{st}$ and the $3^{d}$ and $2^{nd}$ beams. If the centers of the beams are not displaced then Figs. 3c, 4c, 10, 12 show what can happen to focused and non-focused beams. Quasi-periodic structures are formed with vortex singularities and topological charge equal to 1 or -1. If focused beams, with displaced centers, interact with each other then a system of three clockwise-rotating vortices with one counter-clockwise rotating vortex in the center can be formed, as shown in Figs. 5c, 11.

If the absolute values of the beam amplitudes are the equal then the main effect of the nonlinear diffraction is to broaden the interacting structures. For focused beams, a broadening of the region of interaction around the central point also occurs. If the modulus of the amplitudes of interacting beams with shifted centers are different, nonlinearity causes shifting and a rotation of the structure "as a whole". A vortex exists in the linear case, surrounded by three symmetrical vortices. A vortex structure of three beams with different absolute values of the input amplitudes changes qualitatively due to nonlinearity. At the same time, no nonlinear change of "vortex charge" is obtained under the condition of a small level of higher harmonics adopted in the present modeling.

One direction for future investigation is to search for a nonlinear change in the vortex charge using layered structures that include materials with higher nonlinear coefficients such as ferrite-paraelectric-dielectric, or just paraelectric waveguides, where dispersion is smaller and nonlinearity larger than in ferrites. As a result, the effective generation of higher harmonics can be expected, and, therefore, a nonlinear vortex charge change. Another interesting possibility is the generation of vortex structures using an amplitude-dependent group velocity, which leads to the propagation of different parts of a pulse, with different velocities and phase defect structure formation [14]. To do this, structures with large nonlinearity and large enough dispersion are necessary. Probably, paraelectric-ferrite structures [20] combining the high dispersion of ferrite films and the large paraelectric nonlinearity [21] would be suitable for this. The control of vortex structure characteristics could be achieved using two counter-propagating pulses that are in anti-phase and include a relative shifting of their centers, together with parametric coupling between them. This kind of parametric coupling could be provided using a microwave resonator [23]. There are indeed a lot of interesting possibilities for cm and mm range vortex research based upon multilayered structures that include ferrite films.

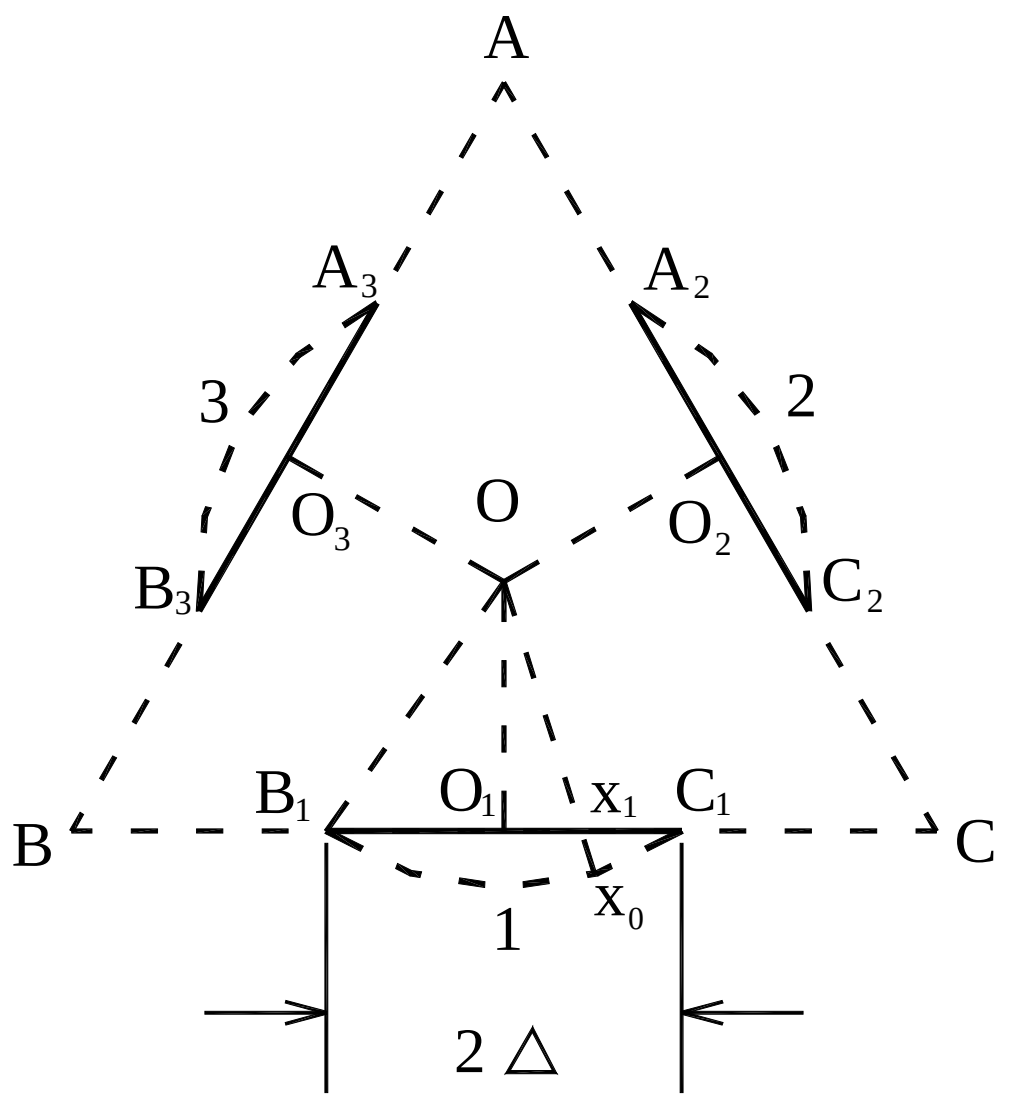

Fig. 2

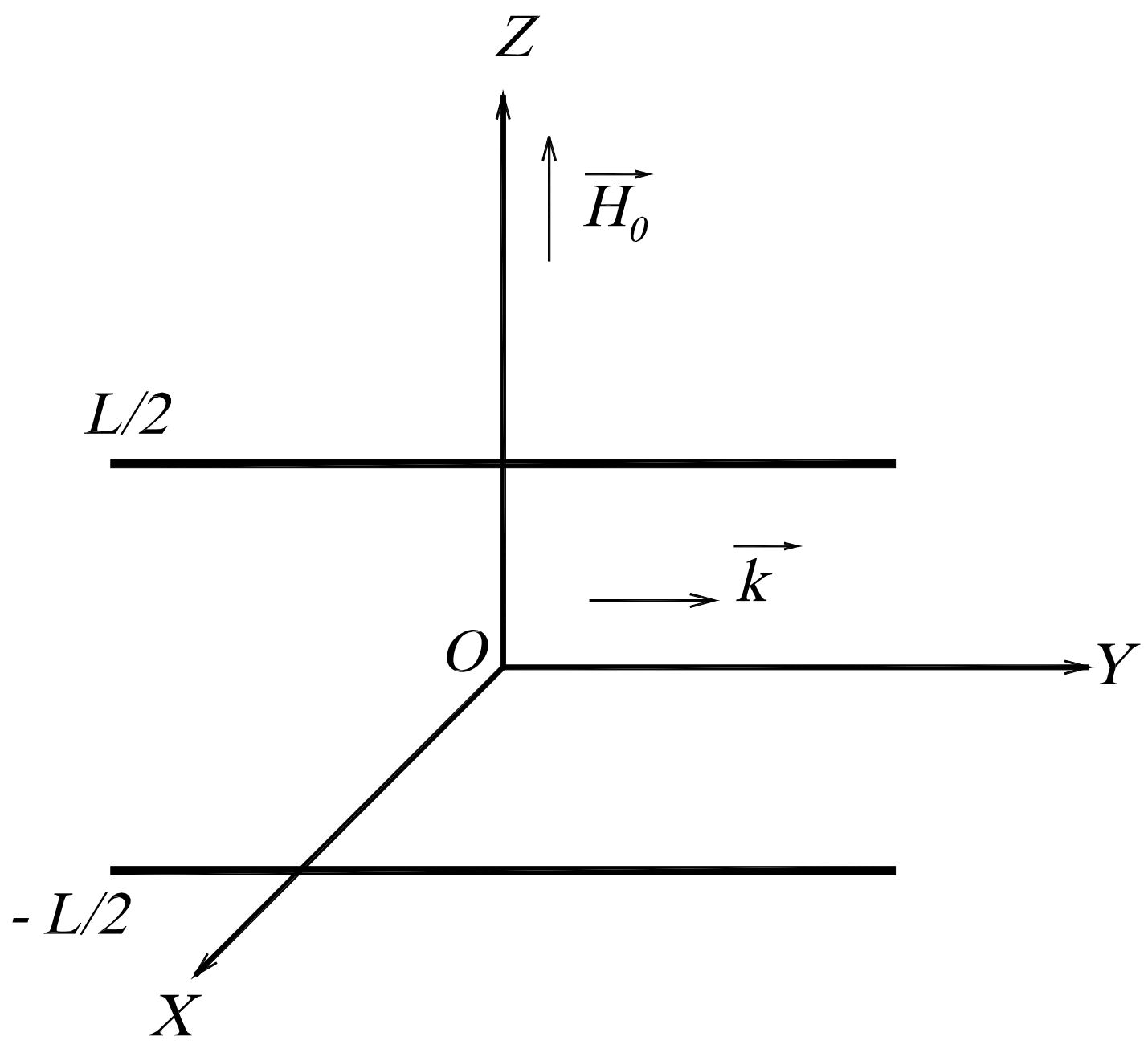

Fig. 1

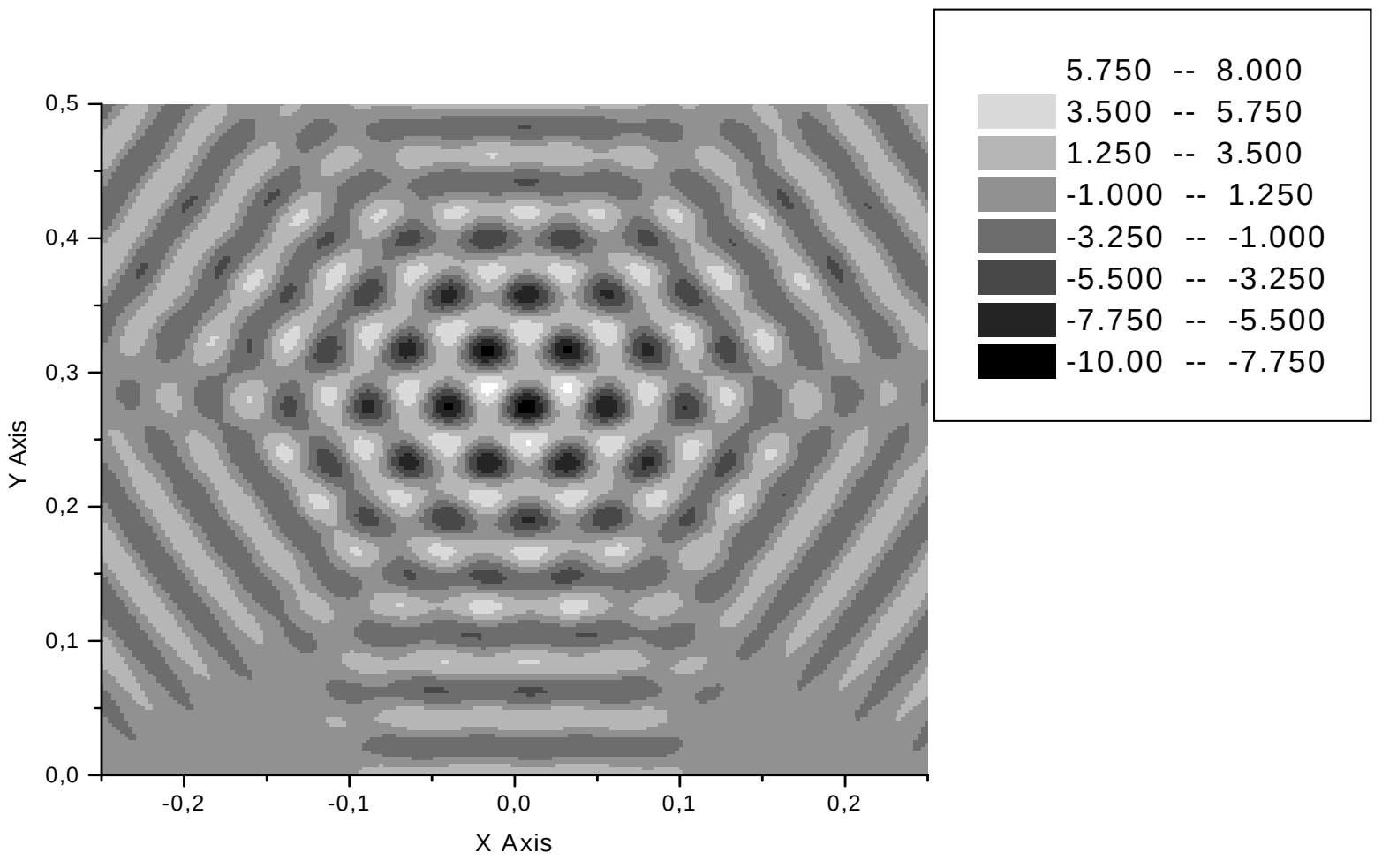

Fig. 3a

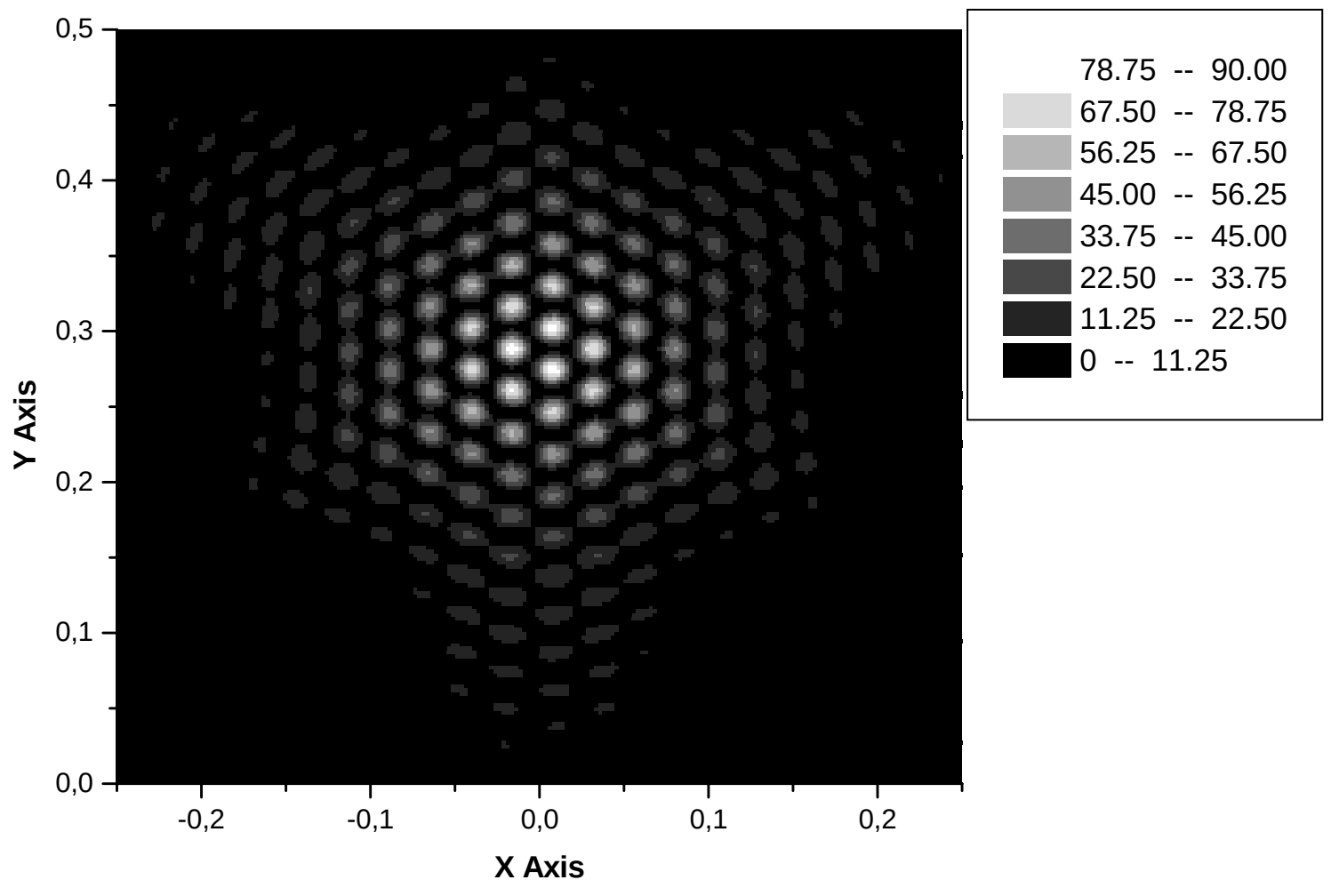

Fig. 3b

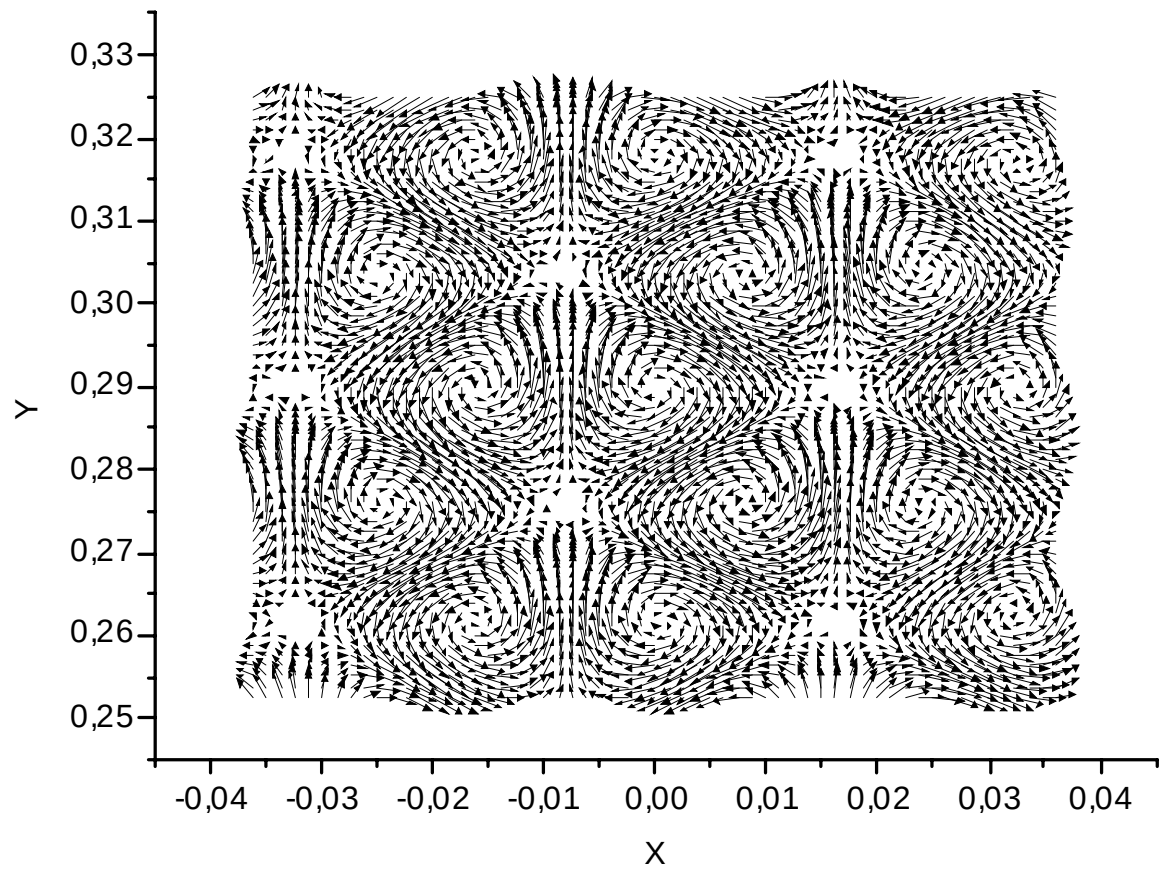

Fig. 3c

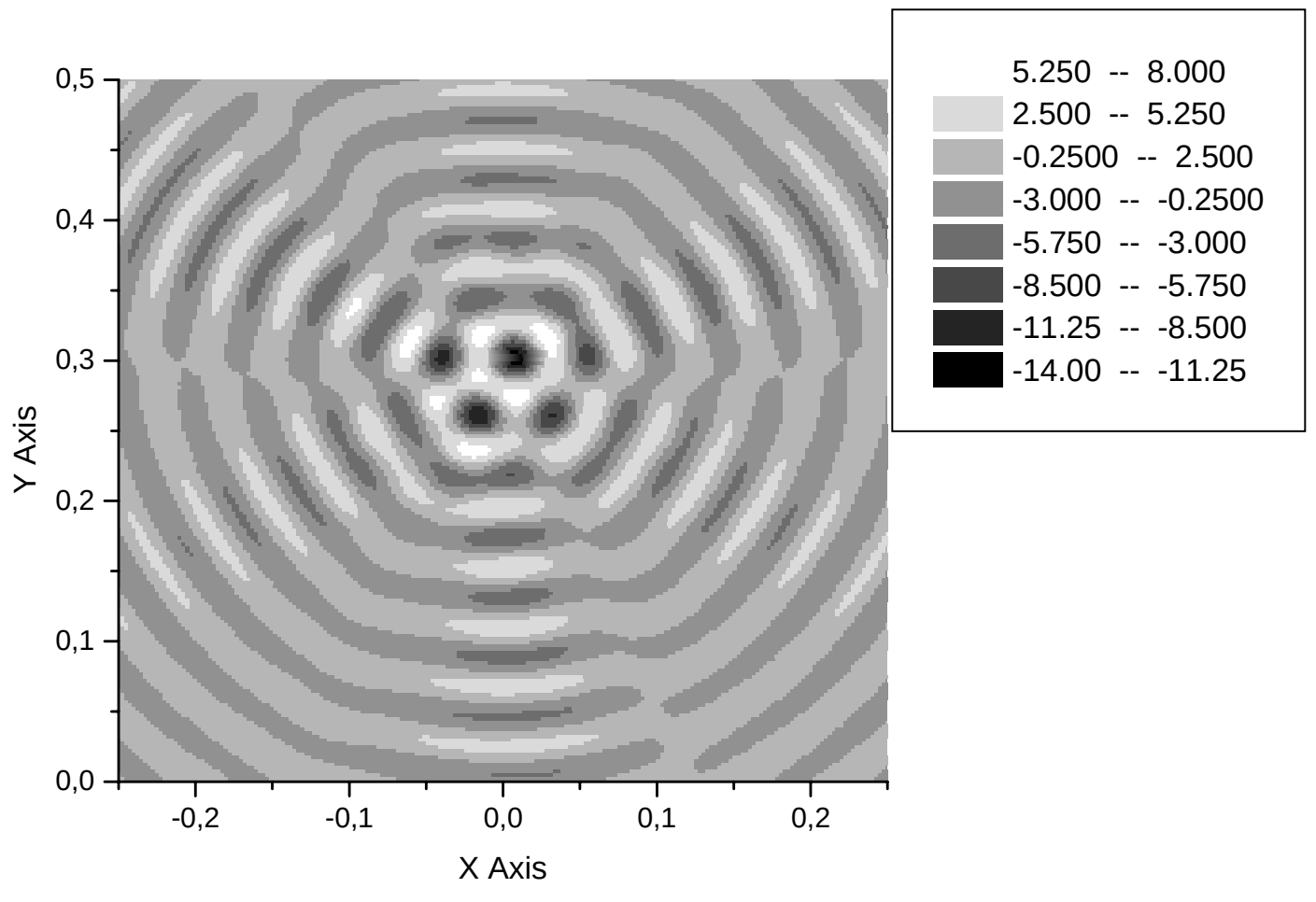

Fig. 4a

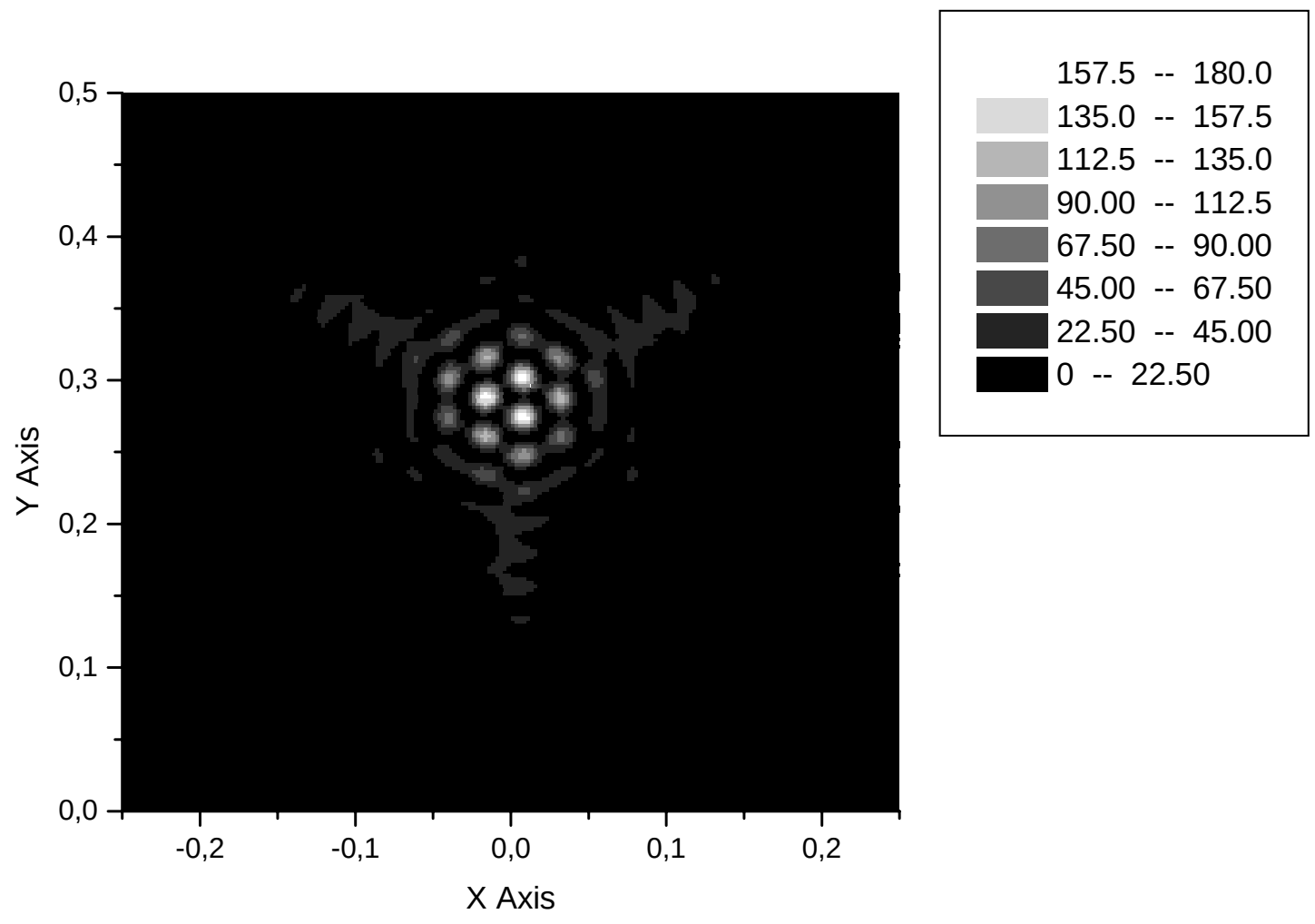

Fig. 4b

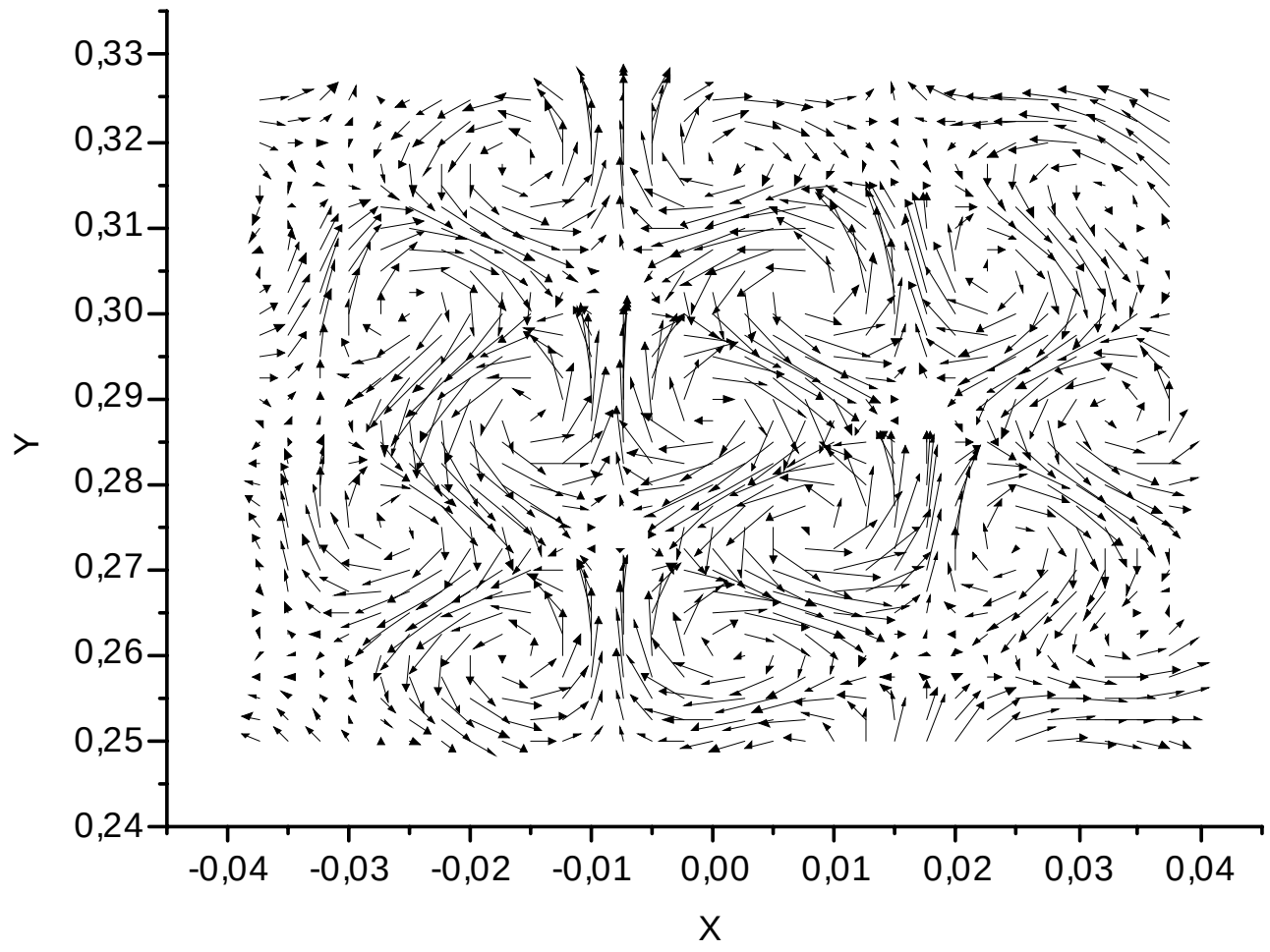

Fig. 4c

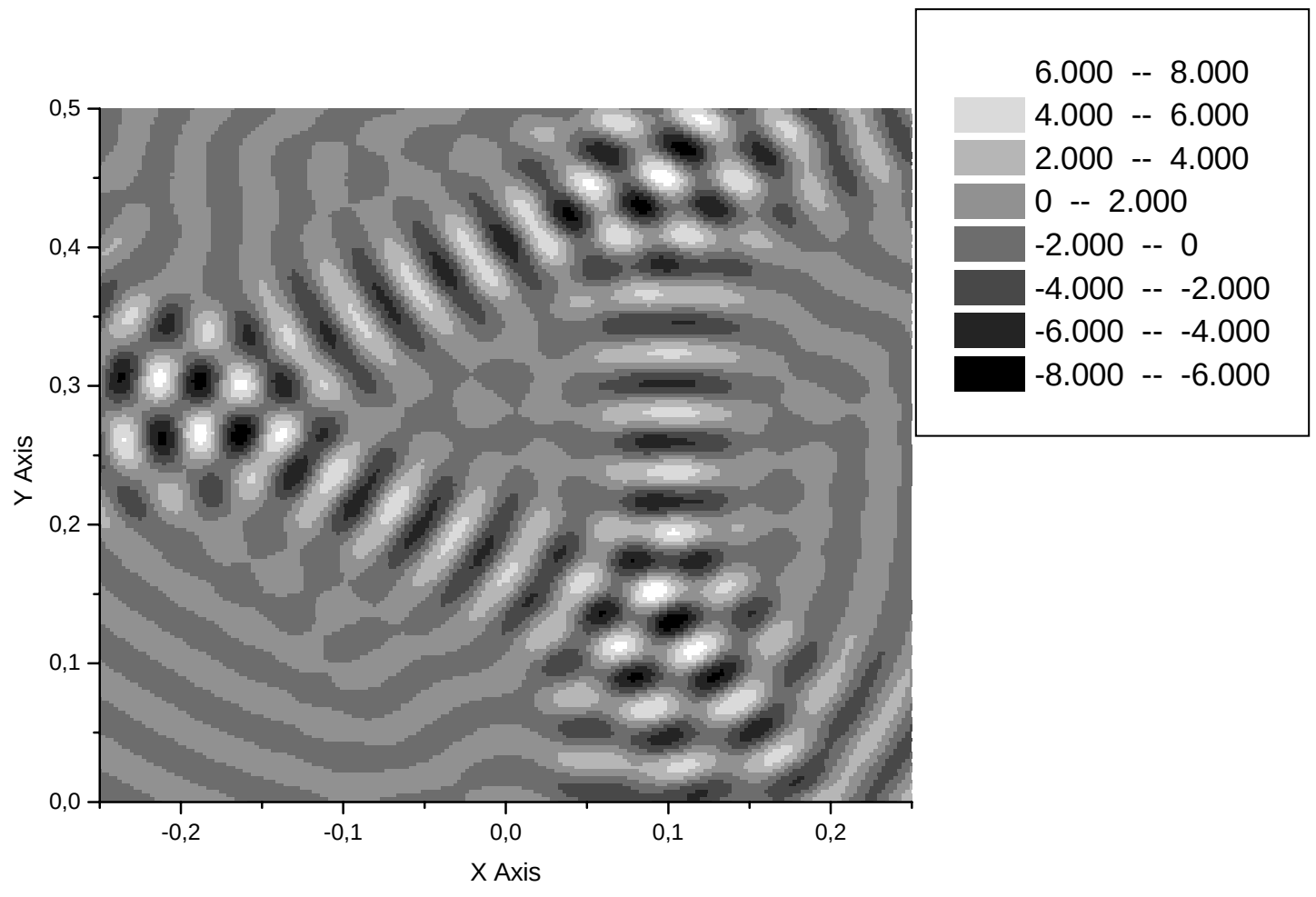

Fig. 5a

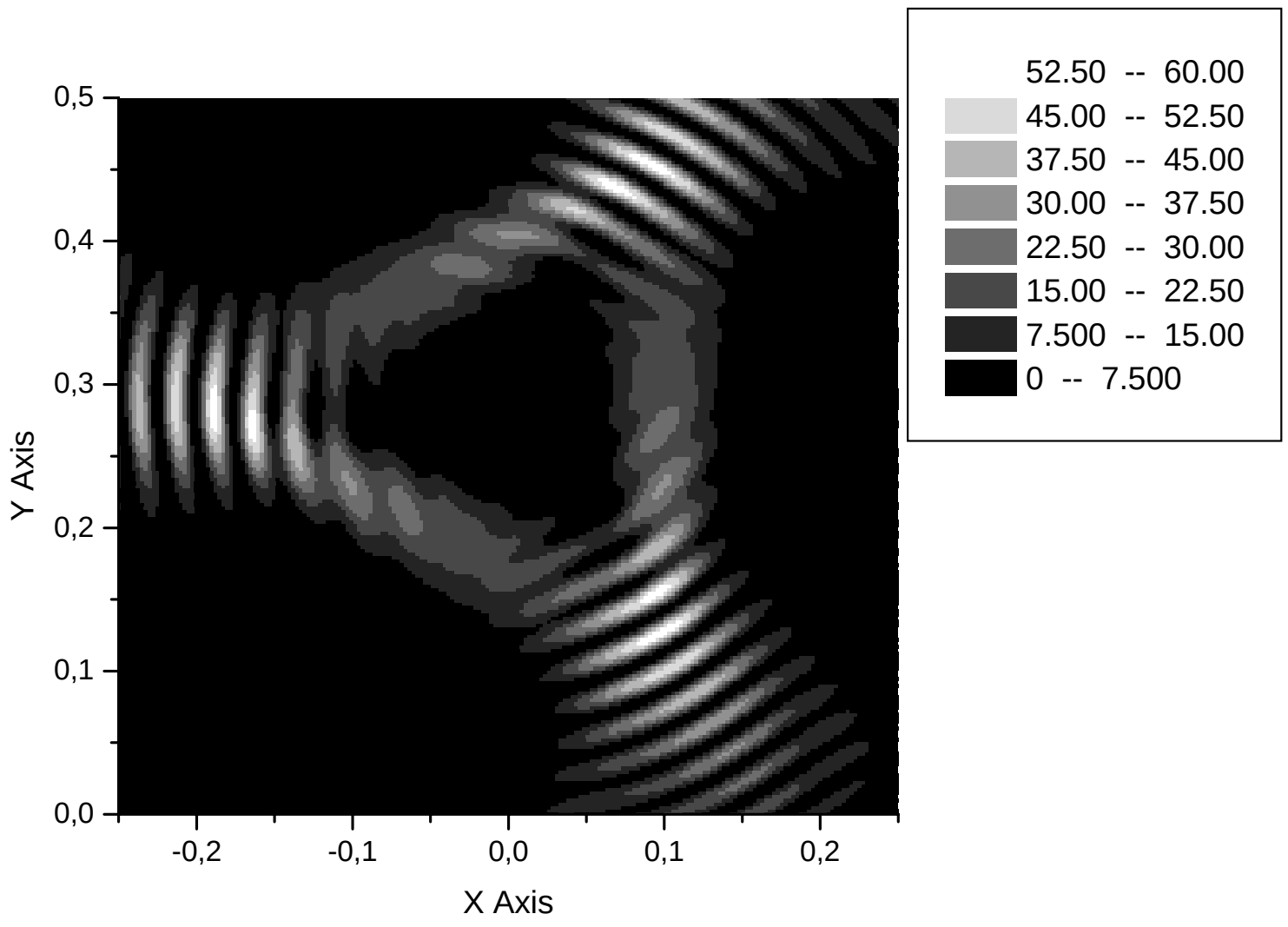

Fig. 5b

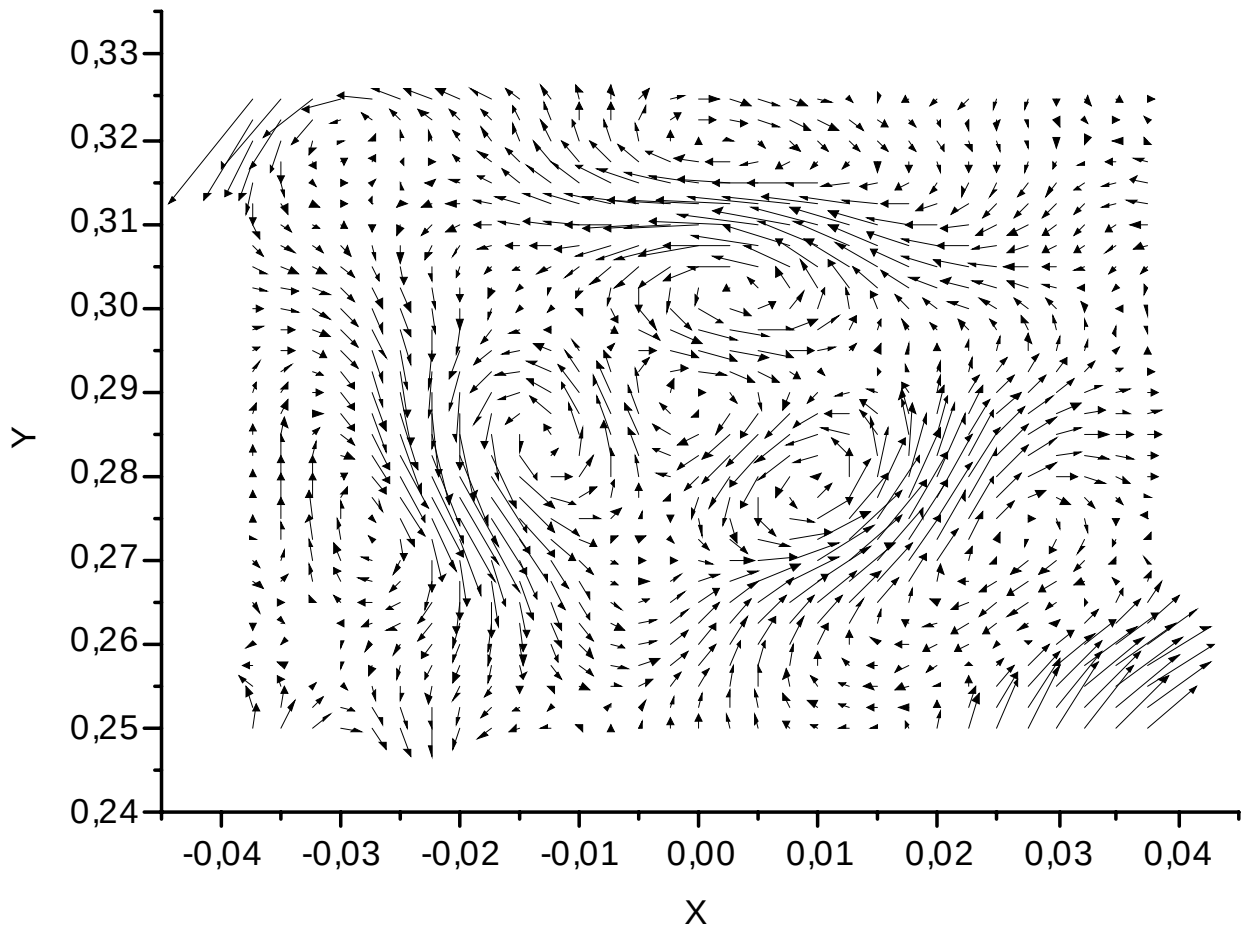

Fig. 5c

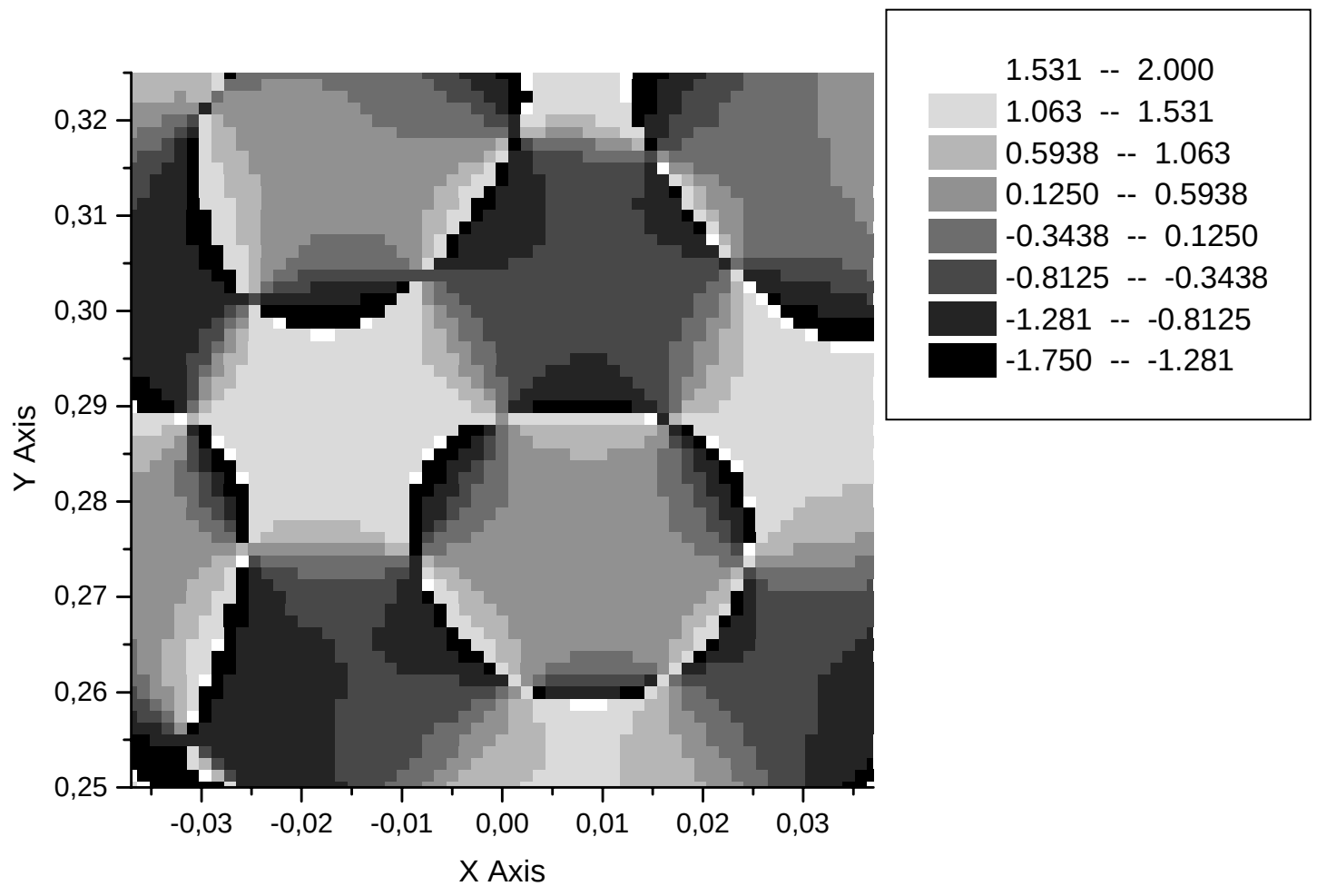

Fig. 6a

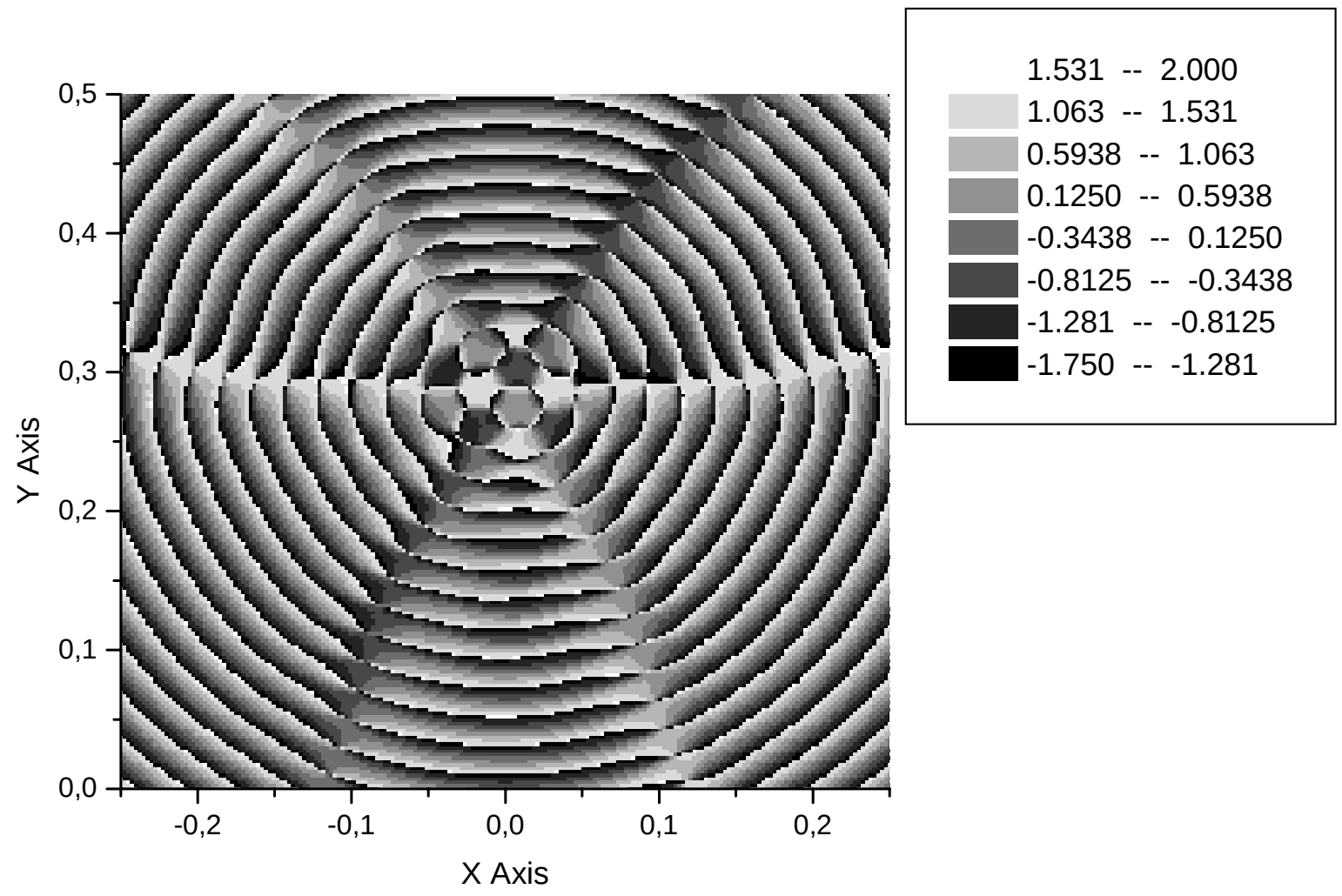

Fig. 6b

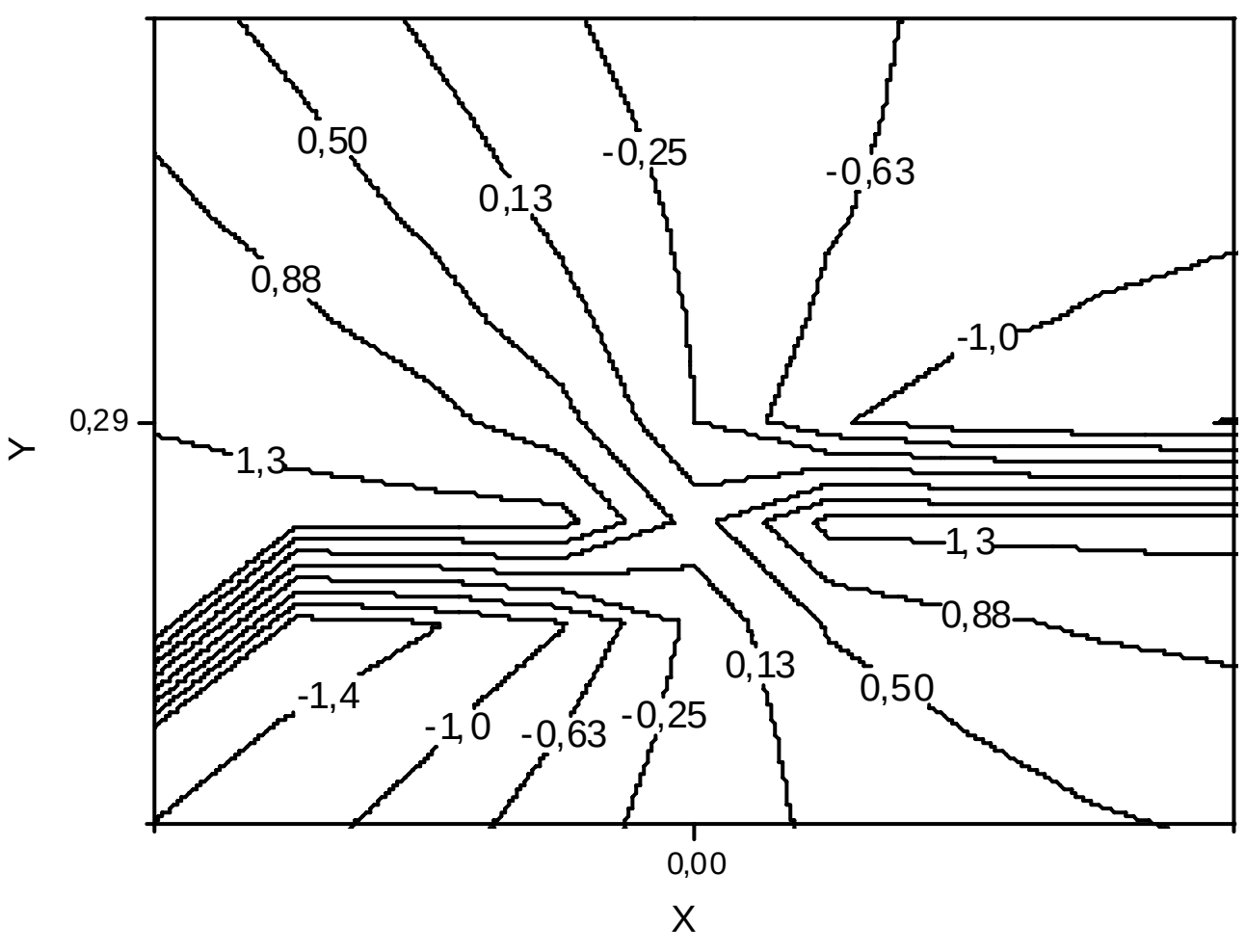

Fig. 7

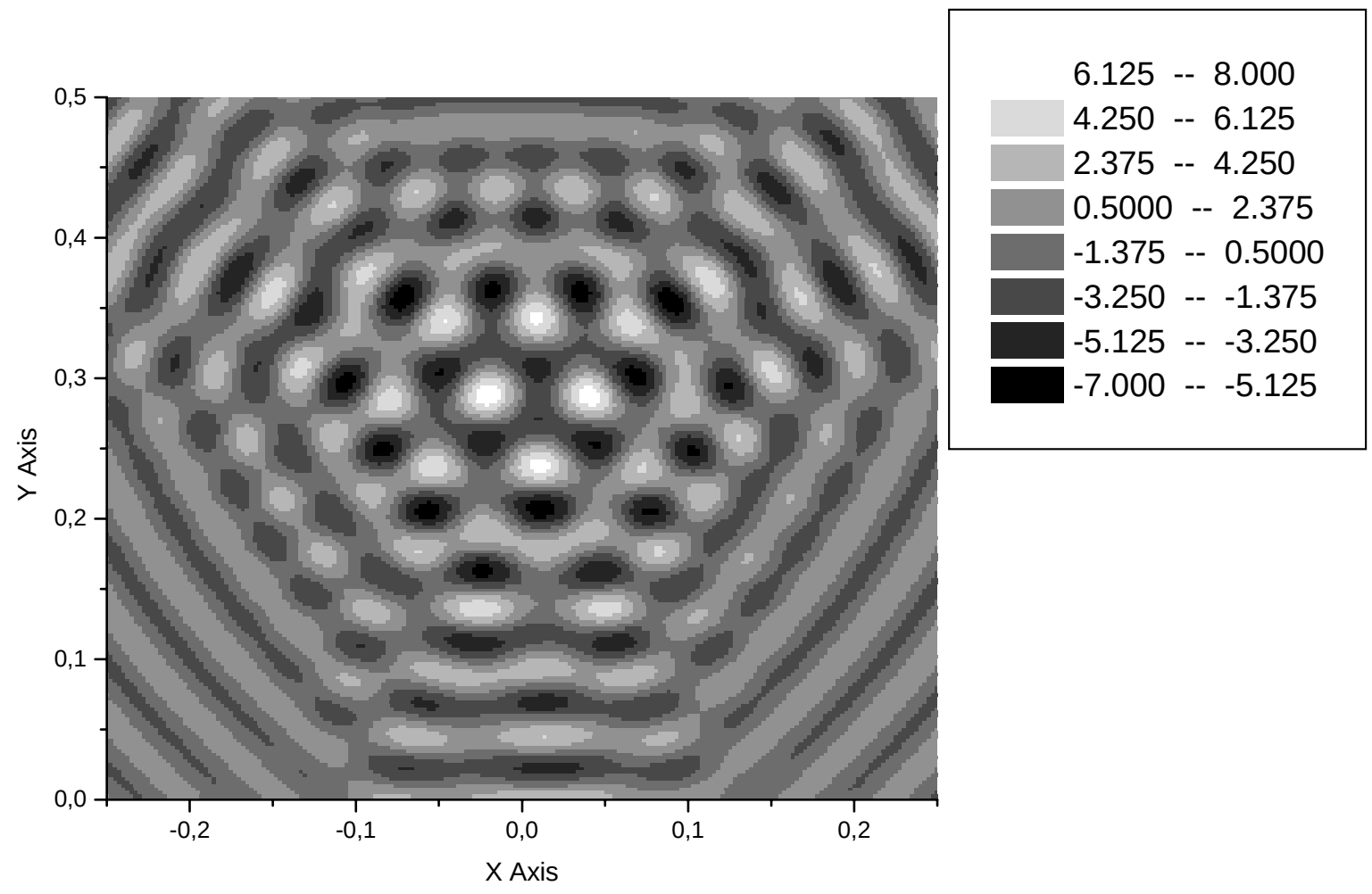

Fig. 8

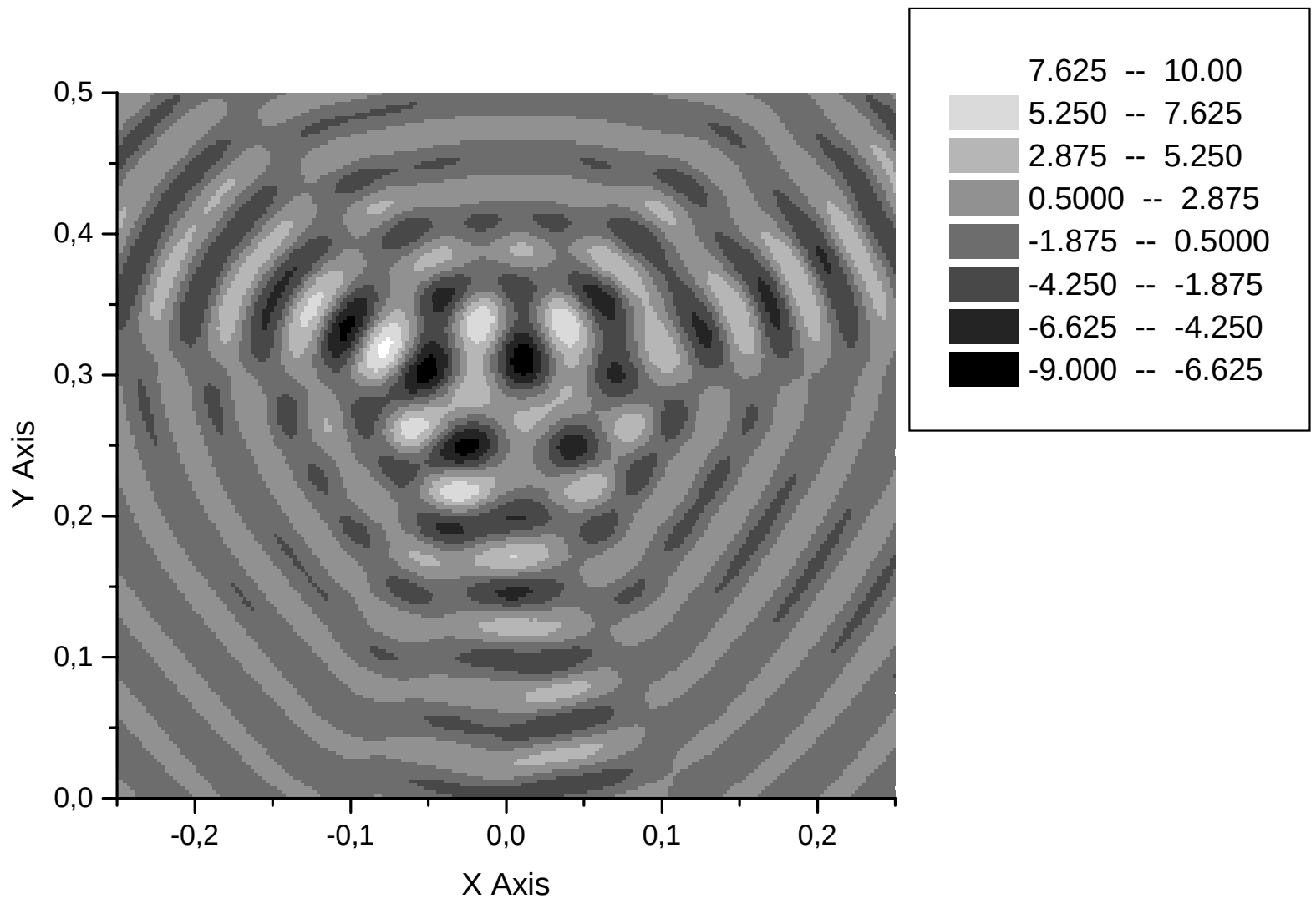

Fig. 9

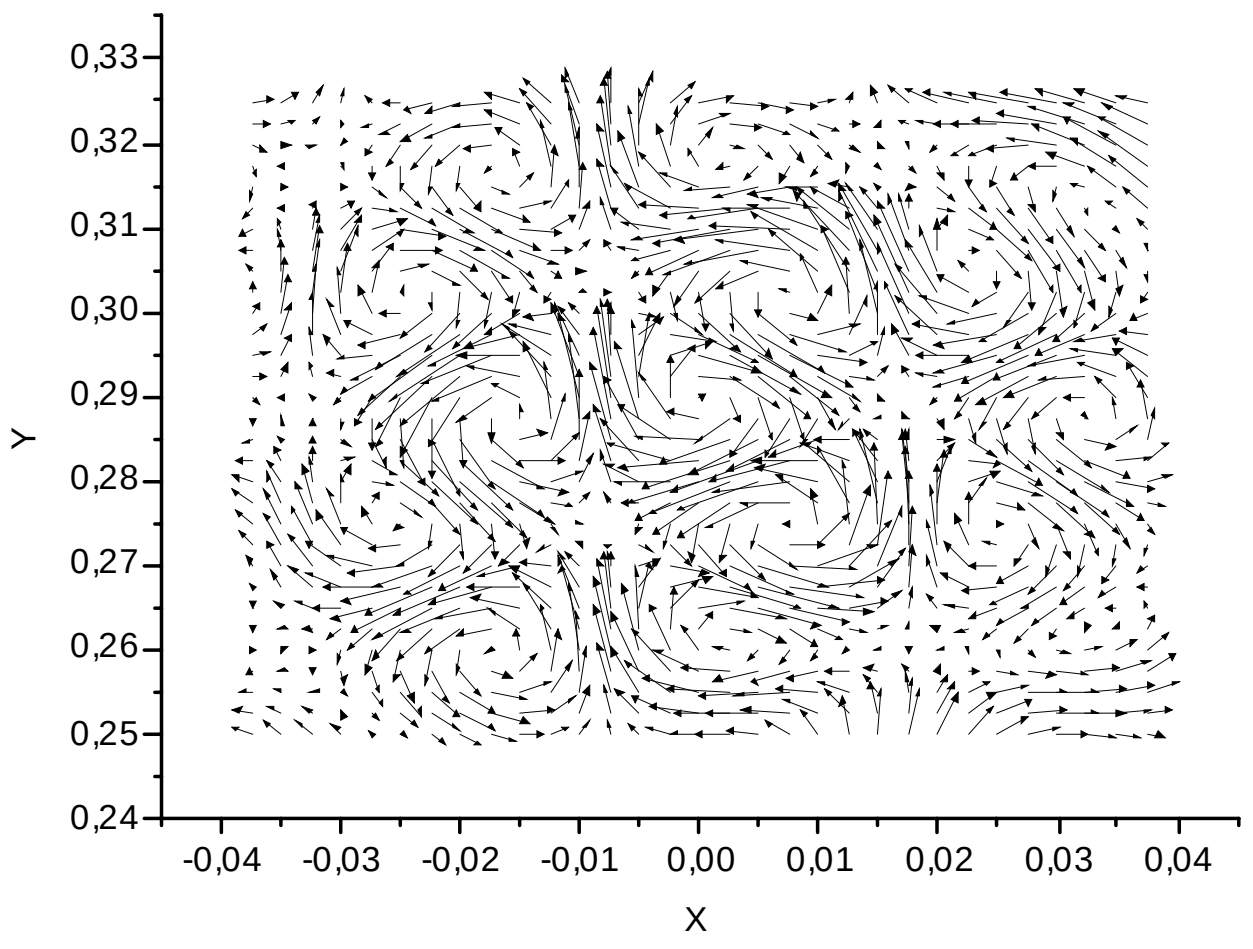

Fig. 10

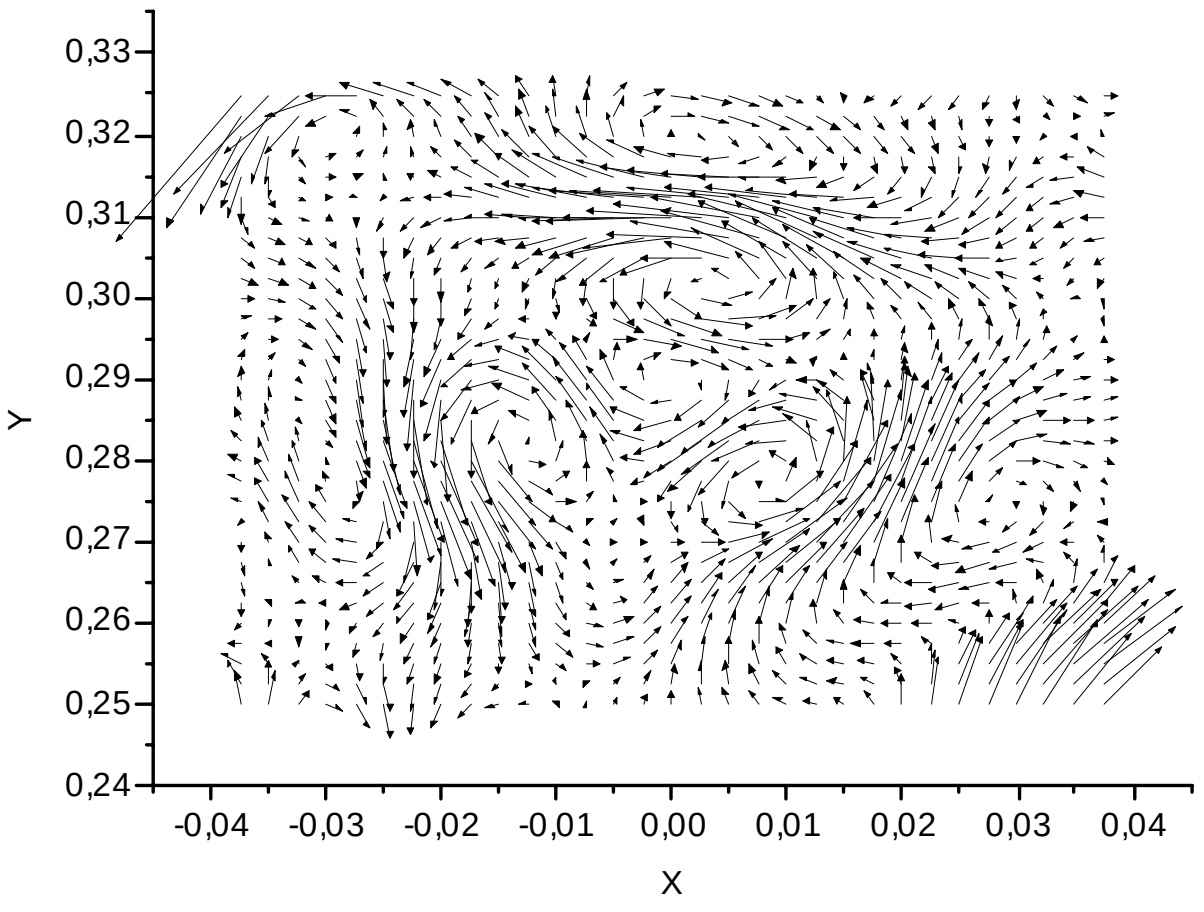

Fig. 11

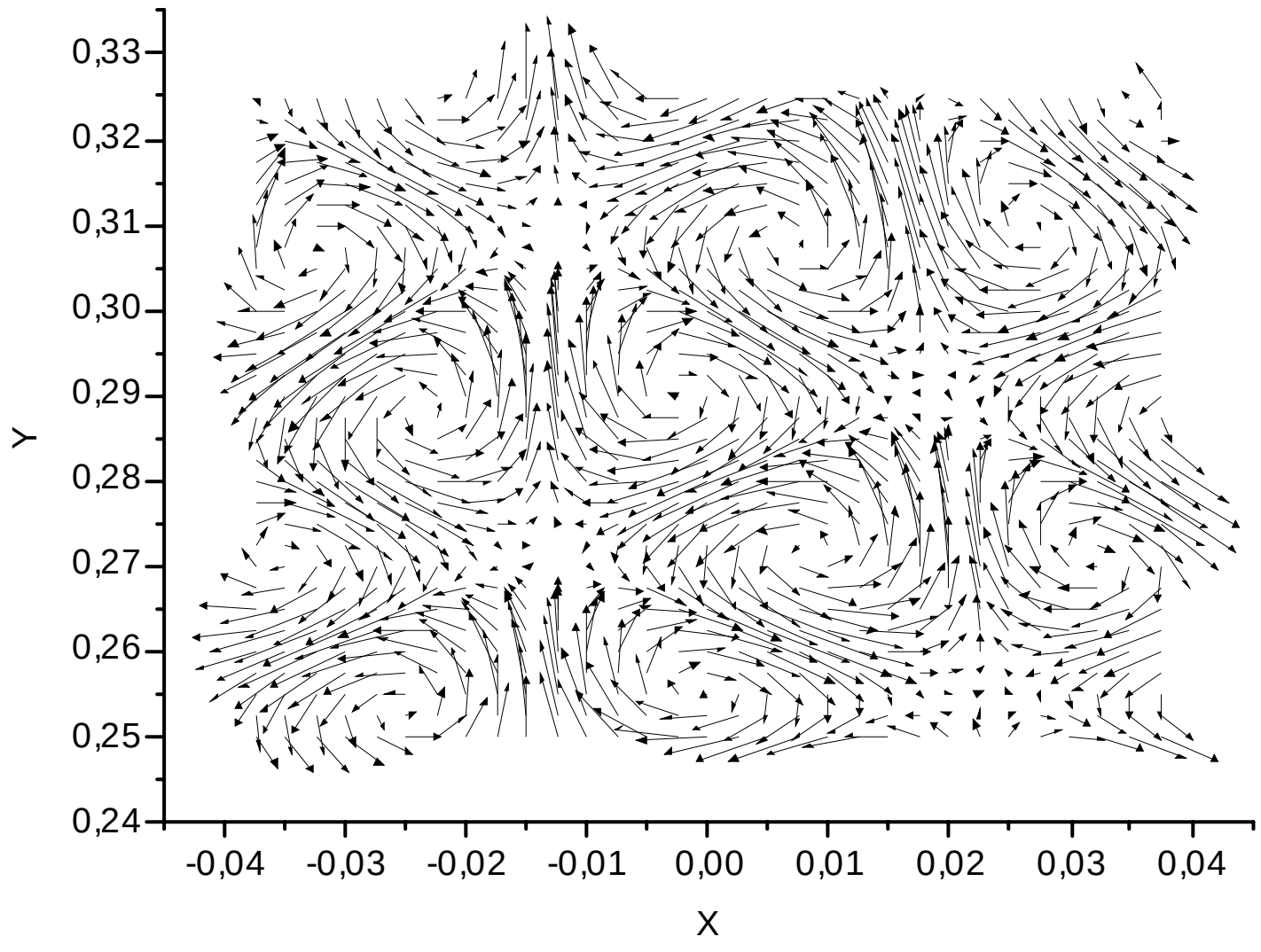

Fig. 12

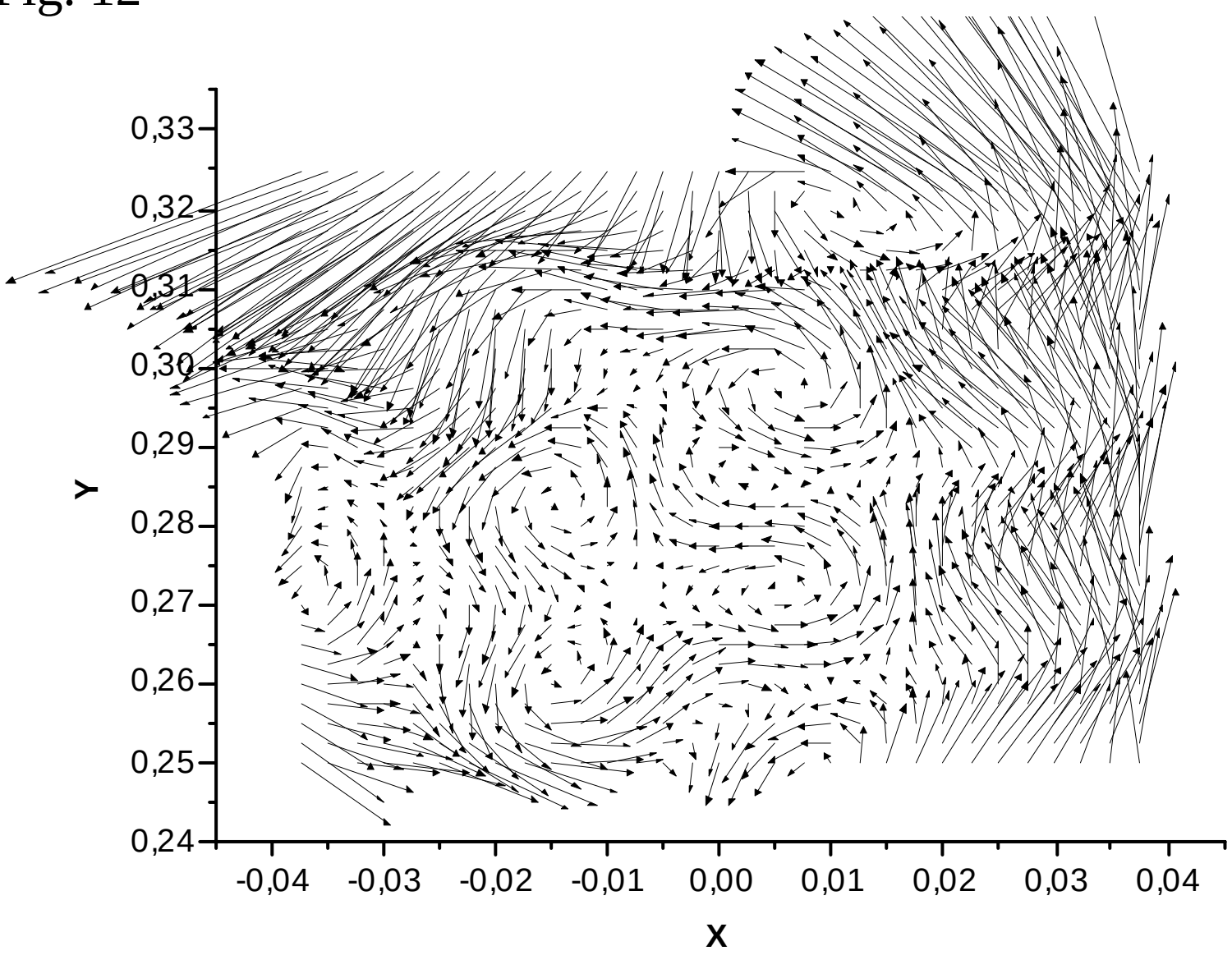

Fig. 13

**Figure Captions**

Fig. 1 Forward volume magnetostatic wave [FVMSW] propagating in normally magnetized ferrite film. L is the width of the film, **k** is the wave number and $\mathbf{H_o}$ is the applied magnetic field

Fig. 2 Excitation of three waves by plane antennas (solid line sections $B_1O_1C_1$, $A_2O_2C_2$ and $A_3O_3B_3$), or by arc antennae (arcs $B_1C_1$, $A_2C_2$ and $A_3B_3$ of a circle around O shown as dashed lines; these arcs are not shown to scale for the convenience of displaying the geometrical calculations). Circular antennae are placed at $\frac{2\pi}{3}$ intervals around the center at O.

Fig. 3 Spatial vortex structure excited by three, linear, equal absolute amplitude, unshifted ($x_{i0} = 0, i = 1,2,3$), un-focused magnetostatic plane wave beams The X,Y axes lie along the directions of the lines $O_1C$ and $O_1O$ in the Fig. 2, respectively. The origin of the coordinates coincides with the point $O_1$ (center of the antenna of the first MSW beam) in the Fig. 2. The same applies to figures 4-13 below; the beam width is $x_0 = 0.1$ and remains at this value for the computations displayed in this paper. (a) spatial distribution of $\mathrm{Re}\,\varphi(x, y)$, the real part of magnetostatic potential

(b) distribution of $|\varphi(x, y)|^2$

(c) display of a vector that is proportional to the Poynting vector.

Fig. 4. Spatial vortex structure excited by three, linear, equal amplitude, unshifted, focused beams. (a) spatial distribution of $\mathrm{Re}\,\varphi(x, y)$;( b) $|\varphi(x, y)|^2$; (c) vector proportional to the Poynting vector.

Fig. 5. Spatial vortex structure excited by three, linear, equal amplitude, shifted ($x_{i0} = 0.1, i = 1,2,3$), focused beams. (a) spatial distribution of $\mathrm{Re}\,\varphi(x, y)$; (b) $|\varphi(x, y)|^2$; (c) Poynting vector distribution. .

Fig. 6. a- Spatial distribution of the magnetostatic potential phase $\Phi$ in the neighbourhood of the center point O, for three, equal amplitude, linear, focused beams; b- Spatial distribution of magnetostatic potential $\Phi$ for the same parameters as Fig. 6a, but in wider spatial region.

Fig. 7. "Contour map" fragment of spatial distribution (shown in Fig. 6) of magnetostatic potential phase, $\Phi$ in the closed vicinity of the point O, geometrical center of the structure (shown in Fig. 2). The values of $\Phi$ in radians on the lines of the constant phase are shown by the numbers inside the figure**.**

Fig. 8 Nonlinear structure $\mathrm{Re}\,\varphi(x, y)$ for the same parameters as Fig. 3a. Dimensionless input beam amplitudes $U_{10} = U_{20} = U_{30} = 3$.

Fig. 9 Nonlinear structure $\mathrm{Re}\,\varphi(x, y)$ for the same parameters as Fig. 4a. Dimensionless input beam amplitudes $U_{10} = U_{20} = U_{30} = 2.95$

Fig. 10 Spatial distribution of field of the Poynting vector excited by three, unshifted, non-equal amplitude, linear, focused beams. Input beam amplitudes are $U_{10} = U_{20} = 2.5, U_{30} = 2.0$.

Fig. 11 Spatial distribution of field of the Poynting vector excited by three, shifted, non-equal amplitude, linear, focused beams. Absolute values of input beam amplitudes are $U_{10} = U_{20} = 2.5, U_{30} = 2$.

Fig. 12 Distribution of Poynting vector for the same parameters as Fig. 10 but now with a nonlinear interaction of the magnetostatic beams.

Fig. 13 Distribution of Poynting vector for the same parameters as Fig. 11, taking into account a nonlinear interaction of the beams.

**References**

[1] V.I.Petviashvili, JETP Letters, **32**, 619 (1980).

[2] P.K.Shukla, Physica Scripta, **32**, 141 (1985).

[3] N. Kukharkin and S. A.Orszag, Phys. Rev. E, **54**, R4524-R4527 (1996).

[4] V.M. Chmyrev, S.V. Bilichenko, O.A. Pokhotelov, V.A. Marchenko, V.I. Lazarev, A.V. Streltsov and L. Stenflo, Phys. Scripta, 38, 841 (1988)

[5] V.A.Andryshenko, Izvestiya AS USSR, Mechanics of Liquid and Gas, **N 2**, 186 (in Russian) (1978).

[6] B.Luther-Davis, J.Christou., V.Tikhonenko and Yu V. Kivshar, J. Opt. Soc. Am. B 1997, **14**, 3045 (1997).

[7] T. Kambe , T. Minota and M.Takaoka, Phys. Rev. E, **48**, 1866 (1993).

[8] J.F.Nye and M.V. Berry, Proc. R. Soc. London A, **336**, 165 (1974).

[9] R.W.Damon and J.R. Eshbach, J. Phys. Chem. Solids, **19**, 308 (1961).

[10] A.I.Ahkiezer, V.G. Bar’yahtar and G., Peletminskiy, Nauka Publ. (1967).

[11] J.Masajada and B.Dubik, Optics Commun., **198**, 21(2001)

[12] A.K. Zvezdin and A.F. Popkov, Sov. Physics JETP), **84**, 606 (1983)

[13] V.V. Grimalsky, Yu G.Rapoport and A.N. Slavin, J. Phys. IV France, **7**, C1-393 (1997).

[14] M. S. Soskin and M.V. Vasnetsov, Pure Appl. Opt. , **7**, 301 (1998).

[15] M.S.Soskin, V.N. Gorshkov, M..V. Vasnetsov, J.T. Malos and N.R. Heckenberg , Phys. Rev. A, **56**, 4064(1997).

[16]. I. V. Basistiy, M. S. Soskin and M. V. Vasnetsov, Opt. Comm.,**119**, 604(1995).

[17] A.G. Gurevich and G.A. Melkov, CRS Press, Boca Raton, 1996.

[18]. H. Suhl, J. Phys. Chem. Solids, **1**, 209 (1957).

[19] Y.T.Zhang, C.E. Patton and G. Srinivasan, J. Appl. Phys., **63**, 5433 (1988).

[20] V.Demidov, P.Edenhofer and B. Kalinikos, Electron. Lett., **37**, 1154 (2001)

[21]. V.V. Grimalsky and S. V. Koshevaya, Letters to Journal of Technical Phys. (in Russian), **13 ,**1070 (1987)

[22] A.V.Bagada, G.A.Melkov, A.A.Serga and A.N. Slavin, Phys. Rev. Lett., **79**, 2137 (1997).

[23] M.. Bauer, O.Buttner, S.O.Demokritov, B.Hillebrands, V. Grimalsky, Y. Rapoport and A.N.Slavin, Phys.Rev.Lett., **81**,3769 (1998).